\documentclass[journal=jctcce,manuscript=article,layout=traditional]{achemso}

\usepackage{amsmath}
\usepackage{amsfonts}
\usepackage{amssymb}
\usepackage{graphicx}
\usepackage{color}
\usepackage{physics}
\usepackage{array}
\usepackage{xr}
\usepackage[toc,page]{appendix}

\makeindex

\newcommand{\Hzero}{\hat{H}_0}

\newcommand{\dk}{\hat{d}_{\mu}\kappa_{\mu}}

\newcommand{\cdkij}[2]{C_{{#1}{#2}}}
\newcommand{\sdkij}[2]{S_{{#1}{#2}}}
\newcommand{\dkijn}[3]{M_{{#1}{#2}}^{({#3})}}
\newcommand{\dkij}[2]{M_{{#1}{#2}}}

\title{ Nonlinear Light Absorption in Many-Electron Systems Excited by an Instantaneous Electric Field: A Non-Perturbative Approach}

\author{Alberto Guandalini}
\affiliation{Dipartimento di Scienze Fisiche, Informatiche e Matematiche, Universit{\`a} di Modena e Reggio Emilia, Via Campi 213A, I-41125 Modena, Italy}
\email{alberto.guandalini@unimore.it}
\alsoaffiliation{CNR -- Istituto Nanoscienze, Via Campi 213A, I-41125 Modena, Italy}
\author{Caterina Cocchi}
\affiliation{Physics Department and IRIS Adlershof, Humboldt-Universit\"at zu Berlin, Zum Gro{\ss}en Windkanal 6, D-12489 Berlin, Germany}
\altaffiliation{Carl von Ossietzky Universit\"at Oldenburg, Physics Department, D-26129 Oldenburg, Germany}
\author{Stefano Pittalis}
\affiliation{CNR -- Istituto Nanoscienze, Via Campi 213A, I-41125 Modena, Italy}
\author{Alice Ruini}
\affiliation{Dipartimento di Scienze Fisiche, Informatiche e Matematiche, Universit{\`a} di Modena e Reggio Emilia, Via Campi 213A, I-41125 Modena, Italy}
\alsoaffiliation{CNR -- Istituto Nanoscienze, Via Campi 213A, I-41125 Modena, Italy}
\author{Carlo Andrea Rozzi}
\affiliation{CNR -- Istituto Nanoscienze, Via Campi 213A, I-41125 Modena, Italy}
\email{carloandrea.rozzi@nano.cnr.it }

\begin{document}

\maketitle

\sloppy

\begin{tocentry}
\includegraphics[scale=0.50]{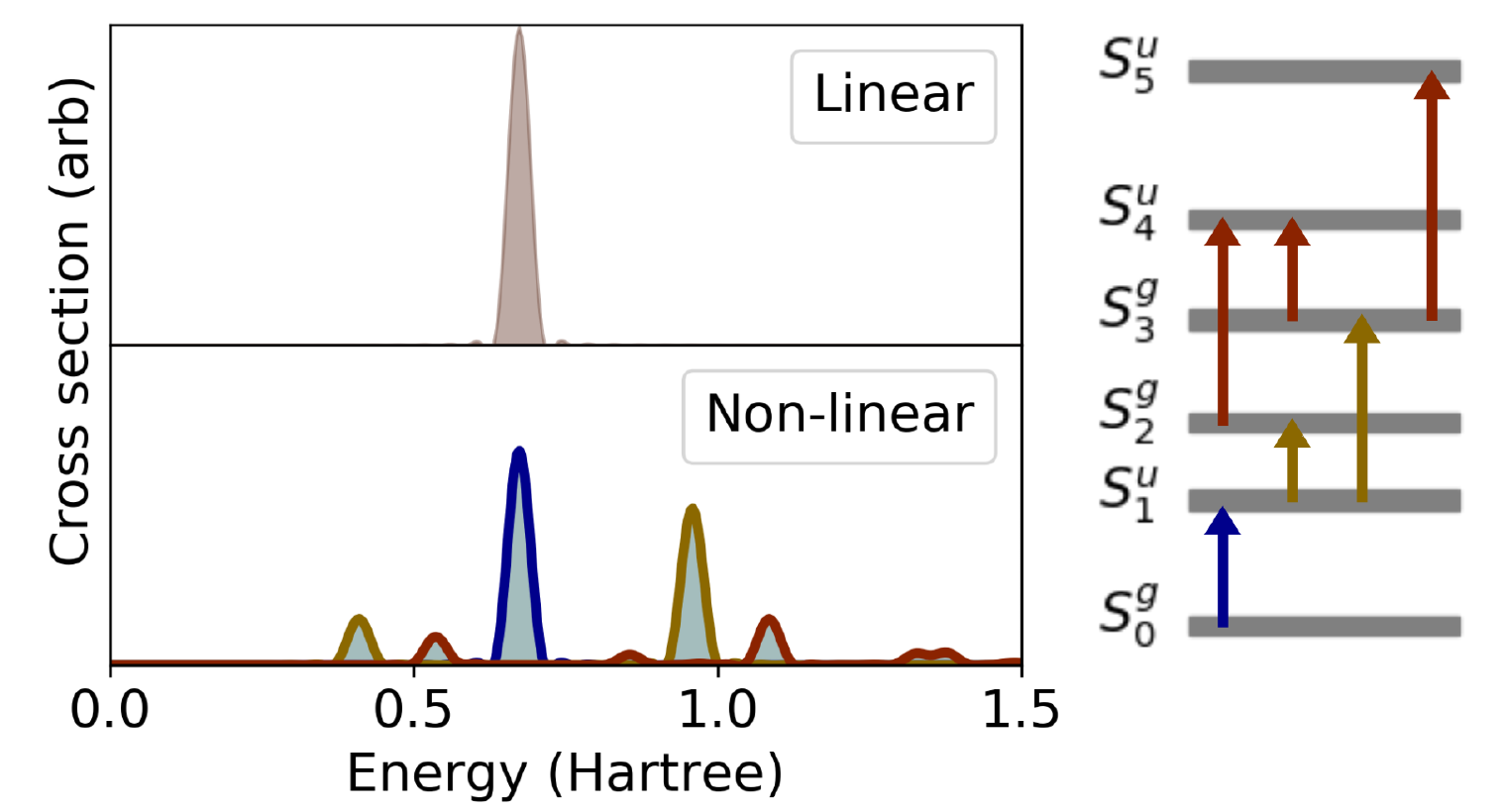}
\end{tocentry}

\newpage
\abstract{
We study light absorption in many-electron interacting systems beyond the linear regime by using a {\em single} broadband impulse of an electric field in the instantaneous limit. We determine non-pertubatively the absorption cross section from the Fourier transform of the time-dependent induced dipole moment, which can be obtained from the time evolution of the wavefunction. We discuss the dependence of the resulting cross section on the magnitude of the impulse and we highlight the advantages of this method in comparison with perturbation theory working on a one-dimensional
model system for which numerically exact solutions are accessible. Thus we demonstrate that the considered non pertubative approach provides us with an effective tool for investigating fluence-dependent nonlinear optical excitations.}

\newpage 
\section{Introduction}
Nonlinear optics globally refers to the regime in which the polarization induced in a material by an electric field is not directly proportional to the magnitude of the external field. All optical media are intrinsically nonlinear, but it is only with the development of high power lasers, that nonlinear properties have become accessible and, hence, extensively studied~\cite{Franken1961,Damm1985,Perry1994}. 

The standard theoretical approach to nonlinear optics rests on perturbation theory, in which the polarization induced in a quantum system by a (classical) electric field is expressed as a power series in the field strength~\cite{Boyd2008}. Susceptibilities calculated at the first few finite orders are usually of interest. Second- and high-order response function theory has been derived in different flavors and levels of accuracy~\cite{Chernyak_2003,Mukamel_2001,Agren_2003,Agren_2005,Gisbergen_1998,Grimme_2018,Haitheinze_2002,Henriksson_2008,Senatore_1986,Yabana_2001,Ye_2006,Kanis_1992,Quinet_2001,Quinet_2001_2,Quinet_2002,Agren_1996,Mukamel_2003,Olsen_1992,Ye_2007,Agren_2002,Agren_2007,Maitra_2011,Parker_2017,Andrade_2007}.
In this way, nonlinear phenomena such as second harmonic generation~\cite{Luppi_2010}, optical rectification~\cite{Hughes_1996,Veithen_2005,Prussel_2018}, and multi-photon absorption in molecular systems~\cite{Friese2015_1,Friese2015_2} can be described.

The development of femtosecond and attosecond lasers has pushed the largest available peak intensity toward magnitudes of the electric field comparable with (or larger than) those experienced by electrons in the atoms~\cite{Brabec2000}. These light sources have thus provided direct access to a variety of resonant regimes in which perturbation theory is not suitable either because the perturbation series does not converge or because its use is impractical~\cite{Lorin2015}. To explore the corresponding phenomenology, a non-perturbative solution is in order~\cite{Peterson1967,Safi2011,Strelkov2016}.

Direct numerical time-propagation of a quantum state subject to time-dependent fields provides us with a numerical approach that, in principle, does not suffer from the limitation of perturbative series expansions~\cite{Goings2018,Isoborn_2016}.
 In fact, nonlinear properties both at finite order in the perturbation~\cite{Cho_2018,Ding_2013,Mattiat_2018} and at all orders~\cite{Bandrauk_2011,Luppi_2012,Nguyen_2016,Tancogne-Dejean_2017,Yabana_2015} can be obtained.
First-principles approaches such as time-dependent density functional theory (TDDFT)~\cite{Ullrich2011,Marques2012} enable realistic simulations of both steady-state and time-resolved spectroscopies for large systems that cannot be tackled by means of  more accurate but also more computationally demanding methods~\cite{Takimoto2007,Attaccalite2013,Uemoto2018,Luppi2012}. The linear and nonlinear regimes can, in principle, be tackled on equal footings within the same framework~\cite{Cocchi2014}. Moreover, the inclusion of nuclear dynamics is straightforward~\cite{Alonso2008, Falke2014,Pittalis2015,Rozzi_2017,Rozzi2018,Yamada2019,Jacobs2020,krumland2020}. Extensions for describing the propagation in presence of decoherence, dissipative environments, or coupling to other external degrees of freedom are also available.~\cite{YuenZhou2010}

When applied to compute the linear-response of a system, the time-propagation method is often formulated in terms of impulse response theory  in order to extract the {\em entire} frequency window of interest by means of a single impulse --- given that the time propagation can be run long enough.  
For time-invariant dynamical systems, the impulse response to an external perturbation is a property of the unperturbed system and is independent of the specific temporal shape of the perturbation: Given the knowledge of the impulse response alone, it is possible to predict the response to any small perturbation by means of the convolution theorem~\cite{Newcomb1963}. In the linear regime, this procedure is equivalent to calculating the first-order polarizability~\cite{Yabana1996}. 

 Although the aforementioned procedure breaks down as the nonlinear effects become important, it  can be extended to compute the nonlinear frequency-dependent cross section in such a way that the determination of the density-density response function does not enter explicitly any step.
As we demonstrate below, the procedure is appealing particularly because it can work well when a perturbative-based approach is challenged by a slow (or a lack of) convergence. The methodology we consider was proposed and applied 
within the framework of real-time TDDFT in the work by Cocchi {\em et al.}~\cite{Cocchi2014} to study optical power limiting due to reverse saturable absorption (RSA) in organic molecules.

RSA is the property of materials to increase their light-absorption efficiency at increasing intensity of the incoming field, due to the presence of an available channel for excited state absorption~\cite{Dini2016,Yaping2019}. It can be described either through phenomenological models~\cite{Miao2019} or in the framework of perturbation theory through a two-step procedure. First, the initial excited state has to be computed and then, from it, the optical absorption spectrum has to be determined~\cite{Grimme_2019,Govind_2015,Govind_2016,Govind_2017,Govind_2019,Maitra_2012,Mosquera_2016,Sheng_2020,Bellier2012,Fischer2016}. However, these methods are limited to one (or at most a few) absorption channels given \textit{a priori}, which hinders the generality of the results.
 In contrast, the real-time methodology by Cocchi {\em et al.}~\cite{Cocchi2014} can capture RSA without any assumption about the excitation channels.
The kind of the nonlinear process that drives the RSA may justify the success of the aforementioned approach: 
As long as the steady-state absorption is observed by means of continuous wave lasers (\textit{i.e.}, we are only concerned with a time-invariant observable) the spectrum depends on the intensity, but not on the detailed shape or phase of the impinging light.

However, for these processes, the interpretation of the results based on TDDFT rests so far on empirical grounds, due to the difficulty of disentangling the approximations introduced by TDDFT from the information provided by the impulsive response. In particular, the adiabatic approximation --- common to essentially all the state-of-the-art TDDFT calculations --- is difficult to improve systematically~\cite{Neepa2014, Neepa2013a, Neepa2013b, Neepa2012, Neepa2005}. Moreover, it is well known that the errors implied by the specific approximations for the ground-state exchange-correlations functional (invoked within in the adiabatic approximation) can vary largely from one form to another~\cite{Head-Gordon2017}. The dependence of the results on these approximations can be expected to be stronger when dealing with non-linear excitations and, thus, further  systematic evaluations would be required.

Here, in order to gain non-empirical insights and provide a sound theoretical foundation to the use of the nonlinear impulse response in nonlinear optics, we focus on a model system for which numerically exact solutions are accessible and, thus, none of the aforementioned approximations is needed. We calculate the dipole moment induced by impulses with increasing strength, and we analyze the resulting nonlinear total cross section. We isolate contributions originating from different sets of ground and excited-state transitions.

This work is organized as follows: In Section~\ref{SecEf} the expression of the absorption cross section and its spectral resolution is derived for an instantaneous impulsive exciting field of arbitrary strength within the dipole approximation. The cross section is analyzed both in the nonlinear case and in the weak-field limit. The complementarity between the proposed methodology and regular perturbative theory is discussed. In Section~\ref{SecEfAppl} the linear and nonlinear absorption of two electrons interacting in a 1D infinite well is simulated by means of an accurate numerical time-propagation scheme. 
The absorption cross section is interpreted, determining the contribution of ground-state and excited-state absorption at different field strengths. Finally, the validity, the usefulness, and the limitations of the proposed methodology to simulate nonlinear properties in complex materials are discussed.

\section{Absorption Cross Section in the instantaneous Impulsive Limit}\label{SecEf}

Let us consider a system of $N$ interacting electrons subject to a time-dependent classical electric field in the dipole approximation,  described by the Hamiltonian $\hat{H}(t) = \hat{H}_0 -\hat{d}_{\mu}\mathcal{E}_{\mu}(t)$ (summation over repeated indices is understood), where $H_0$ includes the ordinary electron-nuclei and electron-electron Coulomb interaction, $\hat{d}_{\mu} = \sum_{i=1}^{N} \hat{r}_{\mu}^i$ is the electric dipole operator with $\hat{r}_{\mu}^i$, and $\mu=x,y,z$ is the $\mu$-th component of the position operator of the $i$-th electron. Spin-orbit coupling is neglected. We aim to describe the light absorption process in the case of an extremely short pulse, \textit{i.e.} in the limit of a Dirac-delta time-dependence
\begin{equation}\label{deltaField}
  \mathcal{E}_{\mu}(t) = \kappa_{\mu}\delta(t).
\end{equation}
The quantity $\kappa_\mu$ is a constant specifying direction and magnitude of the instantaneous electric field~\cite{Yabana1996}. In the dipole approximation it is independent of the spatial coordinates. We make no particular assumptions about the value of $\kappa \equiv \sqrt{\kappa_{\mu}\kappa_{\mu}}$. Here, we focus our analysis on absorption at equilibrium, \textit{i.e.}, we suppose that the system in its ground state at $t<0$, namely $\ket{\Psi(t<0)} = \ket{\Psi_0} e^{-iE_0 t}.$ 
For a description of the out-of-equilibrium absorption processes, as in the case of time-resolved absorption spectroscopy, further analysis is required~\cite{Perfetto2015}.

For $t > 0$, the solution of the time-dependent Schr\"odinger equation $\ket{\Psi(t)}$ can be projected onto the eigenstates $\left\lbrace \ket{\Psi_i} \right\rbrace$ of $\hat{H}_0$ as
\begin{equation}\label{WFdt}
\ket{\Psi(t>0)} = \sum_{i=0}^{+\infty} c_i \ket{\Psi_i}e^{-iE_it} \ ,
\end{equation}
where $\lbrace E_i\rbrace$ are the eigenvalues of $\hat{H}_0$. The coefficients $c_i$ are
\begin{equation}\label{WFcoeff}
c_i = \mel{\Psi_i}{e^{-i\dk}}{\Psi_0}\ .
\end{equation}
Due to the instantaneous nature of the perturbation, the $c_i$ coefficients are time-independent. The time-dependent dipole moment $d_{\mu}(t) = \mel{\Psi(t)}{\hat{d}_{\mu}}{\Psi(t)}$ for such a system is
\begin{equation}\label{dipt}
d_{\mu}(t) = \theta(t)\sum_{i,j=0}^{+\infty}c_i^*c_j d_{\mu}^{ij}e^{-i\omega_{ji}t}+\theta(-t)d_{\mu}^{00} \ ,
\end{equation}
where $\theta(t)$ is the Heaviside theta function, $d_{\mu}^{ij} = \mel{\Psi_i}{\hat{d}_{\mu}}{\Psi_j}$ are the dipole matrix elements,  $d_{\mu}^{00}$ is the ground state dipole, and $\omega_{ji}=E_{j}-E_{i}$ is the energy difference between the $j$-th and $i$-th eigenstates of the unperturbed Hamiltonian $\hat H_0$.

We now want to employ the explicit shape for the dipole moment in Eq.~\eqref{dipt} in order to calculate the absorption cross section
\begin{equation}\label{DefCrSec}
\sigma(\omega) = \frac{4\pi\omega}{c}\frac{\Im\left[ d_{\mu}(\omega) {\mathcal{E}}^{*}_{\mu}(\omega)\right]}{|{\mathcal{E}}(\omega)|^2} \ .
\end{equation}
The full derivation of Eq.~\eqref{DefCrSec} is provided in the Supporting Infomation, showing that this expression is not limited to the weak-field regime.

By Fourier transforming Eq.~\eqref{dipt}, substituting it into Eq.~\eqref{DefCrSec}, and assuming, for sake of simplicity, that the system is centrosymmetric (\textit{i.e.}, $d_{\mu}^{00} = 0$), we obtain
\begin{equation}\label{CrossSectionEf}
\sigma(\omega) = \frac{4\pi^2}{c\kappa^2}\sum_{i,j>i}
\left[\cdkij{0}{i}\sdkij{j}{0}-\sdkij{0}{i}\cdkij{j}{0}\right] \dkij{i}{j}\omega_{ji} \delta(\omega_{ji}-\omega) \, , 
\end{equation}
where $\cdkij{i}{j} \equiv \mel{\Psi_i}{\cos( \dk )}{\Psi_j}$, $\sdkij{i}{j} \equiv \mel{\Psi_i}{\sin( \dk )}{\Psi_j}$, and $\dkijn{i}{j}{n} \equiv \mel{\Psi_i}{(\dk)^n}{\Psi_j}$, with $M^{(1)}_{ij} \equiv M_{ij} $. Eq.~\eqref{CrossSectionEf}, derived in the Supporting Information, can be used to describe any centrosymmetric system, or an ensemble of non-centrosymmetric objects randomly oriented with respect to the direction of the electric field.
Since the ensemble-averaged optical response is centrosymmetric regardless of the point symmetry of the individual molecules, the average cross section is obtained by sampling and averaging the responses to impulses with different polarization directions.

In order to highlight the nonlinear character of the cross section, we compute the linear absorption cross section $\sigma^{(1)}(\omega)$ by approximating the matrix elements in Eq.~\eqref{CrossSectionEf} up to the first order in $\kappa$. 
We can physically define the concept of ``small $\kappa$" since the magnitude of the dipole is limited by the spatial extension $R$ of the system. For molecular systems with $R\approx (1-100)$ bohr, $\dk << 1$ when $\kappa \ll 1/R$. Therefore, $\kappa \approx 10^{-3} \,\textrm{bohr}^{-1}$ is suitable to define a weak field in the impulsive limit~\cite{Yabana1996}.
Since $\sin(\dk) \approx \dk$ and $\cos(\dk) \approx \hat 1$, we have
\begin{equation}\label{CrossSecLinEf}
\sigma^{(1)}(\omega) = \frac{4\pi^2\omega}{c\kappa^2}\sum_{j=0}^{+\infty}
\left|\dkij{j}{0}\right|^2 \delta(\omega-\omega_{j0})\ .
\end{equation}
For small $\kappa$ (in the above-mentioned sense), Eq.~\eqref{CrossSecLinEf} provides a good approximation for Eq.~\eqref{CrossSectionEf}.


\begin{figure}[ht]
\begin{center}
\includegraphics[scale=0.75]{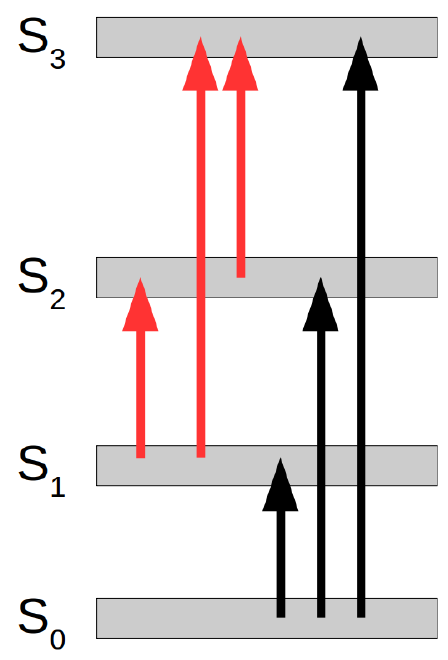}
\caption{Sketch of ground-state (black) and excited-state (red) excitations in the singlet manifold. Excited-state excitations involve transitions between excited states and are activated in the nonlinear regime described by Eq. (\ref{CrossSectionEf}). Ground-state absorption is described by the ordinary linear cross section in Eq. (\ref{CrossSecLinEf}).}
\label{ImageEf+Sp}
\end{center}
\end{figure}

While in the linear regime the resonances in the cross section are only found at $\omega=\omega_{j0} \equiv E_j - E_0$,  in  Eq.~\eqref{CrossSectionEf}
the resonances occur at $\omega=\omega_{ji}  \equiv E_j - E_i$; \textit{i.e.},  for the energy of the incoming light equating the energy difference between any pair of eigenstates of the unperturbed Hamiltonian $\hat{H}_0$. This is sketched in Figure~\ref{ImageEf+Sp}, in which black arrows indicate ground-state absorption (GSA), as given by Eq.~\eqref{CrossSecLinEf}, and red arrows denote excited-state absorption (ESA), as given by Eq.~\eqref{CrossSectionEf}. ESA, in practice, can occur when one or more excited states of the unperturbed system are populated by a laser.

Eqs.~\eqref{CrossSectionEf} and \eqref{CrossSecLinEf} share the same dipole parity selection rule, as they include the same matrix elements $\dkij{i}{j}$, with $i=0$ in Eq.~\eqref{CrossSecLinEf}. However, while states with the same parity as the ground state cannot be populated in the linear regime, they can be populated in the nonlinear regime through ESA. Observing closely, the individual contributions to Eq.~\eqref{CrossSectionEf} vanish  for all the states $i$ with either $\cdkij{0}{i} = 0$, or $\sdkij{0}{i} = 0$. In the next section, we  exploit this  fact to determine the set of transitions that  build up the ESA at different field strengths. Also note that in Eq.~\eqref{CrossSectionEf} $\sigma (\omega)$ satisfies the Thomas--Reiche--Kuhn sum rule in the form
\begin{equation}\label{EfSumRule}
\int\limits_{0}^{+\infty} d\omega \sigma(\omega) = \frac{2\pi^2N}{c} \ ,
\end{equation}
where $N$ is the number of electrons. Since the right hand side of Eq.~\eqref{EfSumRule} does not depend on $\kappa$, the sum rule is valid both in the linear and nonlinear regimes.

The perturbative analysis can be carried on by further expanding Eq.~\eqref{CrossSectionEf} in powers of $\kappa$. The second-order term, like all subsequent terms of even order, vanishes because we assumed inversion symmetry. The third-order term is
\begin{equation}\label{sigma3impulse}
\sigma^{(3)}(\omega) = \frac{4\pi^2}{c\kappa^2} \left[ \sigma^{(3)}_{\rm GSA}(\omega) + \sigma^{(3)}_{\rm ESA}(\omega) \right] \, ,
\end{equation}
where
\begin{equation}\label{sigma3gsa}
\sigma^{(3)}_{\rm GSA}(\omega) \equiv
-\frac{1}{6}\sum_{j}
\left(\dkij{0}{j}
      \dkijn{j}{0}{3}
      +3\dkijn{0}{0}{2}|\dkij{j}{0}|^2
\right)
\omega_{j0}\delta(\omega-\omega_{j0})
\end{equation}
has poles only at the ground state excitation energies $\omega_{j0} = E_{j}-E_{0}$ (see Eq. \ref{CrossSecLinEf}): \textit{i.e.}, it describes the third-order correction to GSA. Hence, the term
\begin{equation}\label{sigma3esa}
\sigma^{(3)}_{\rm ESA}(\omega) \equiv 
-\frac{1}{2}\sum_{\substack{i > 0\\j > i}}
\left( \dkijn{0}{i}{2}\dkij{j}{0}
      -\dkij{0}{i}\dkijn{j}{0}{2}
  \right)
\dkij{i}{j}\omega_{ji}\delta(\omega-\omega_{ji})
\end{equation}
describes the third-order correction to ESA. Note that $\sigma^{(3)}$ includes both excitations and de-excitations from the excited states. The spectral features obtained from the nonlinear cross section may therefore have either positive or negative oscillator strengths, physically corresponding to light absorption or emission from an excited state. Next, the fifth-order correction is derived in a similar way as done for the third-order one:
\begin{align}\label{sigma5imp}
&\sigma^{(5)}(\omega) = \frac{4\pi^2\omega}{c\kappa^2}
\left\lbrace 
\frac{1}{120}\sum_{j}
\dkijn{j}{0}{5}
\dkij{0}{j}\omega_{j0}\delta(\omega-\omega_{j0})
\right. \nonumber \\
&+\frac{1}{12}\sum_{i,j > i}
\left[
  \dkijn{0}{i}{2}\dkijn{j}{0}{3} - \dkijn{0}{i}{3}\dkijn{j}{0}{2}
\right]
\dkij{i}{j}\omega_{ji}\delta(\omega-\omega_{ji}) \nonumber \\
&+\left.\frac{1}{24}\sum_{i,j > i}
\left[
  \dkijn{0}{i}{4}\dkij{j}{0} - \dkij{0}{i}\dkijn{j}{0}{4}
\right]
\dkij{i}{j}\omega_{ji}\delta(\omega-\omega_{ji})
\right\rbrace\ .
\end{align}
GSA and ESA contributions can be identified for $\sigma^{(5)}$ similarly as done for $\sigma^{(3)}$, by isolating the terms with $i=0$ in the latter expression (see Sec.~\ref{Secsigma5Appx} in the Supplementary Information). We will make use of Eq.~\eqref{sigma5imp} in the next section.

Even if Eq.~\eqref{sigma3impulse} also accounts for two-photon processes, the impulsive field has a fixed frequency dependence. Therefore, it cannot predict spectra obtained by means of non-impulsive field shapes. On the other hand, Eq.~\eqref{sigma3impulse} allows us to directly identify the spectral weight due to specific set of transitions at a common resonance. Consequently, accessing $\sigma(\omega)$ with an instantaneous impulse is most useful to describe ESA (which is fluence dependent) but it is not suitable for a full description of two-photon absorption (which is irradiance dependent) \cite{Santhi2006}.

Obviously, the impulse response obtained from real-time propagation of the quantum state is intrinsically non perturbative: namely, from the evolved quantum state we compute $d_\mu(t) = \mel{\Psi(t)}{\hat{d}_\mu}{\Psi(t)}$ and, thus, the Fourier transform $d_\mu(\omega)$ can be readily obtained. The latter can be finally used in Eq.~\eqref{DefCrSec}. According to the previous analysis, the approach captures ESA at {\em all} the possible resonances.
Furthermore, because $d_\mu(t)$ can be expressed in terms of the particle density, $d_\mu(t) = \int d^3r~ r_\mu n({\bf r}, t)$, the procedure based on real-time propagation can be readily implemented in any code that solves the time-dependent Kohn-Sham equations  without the need for the explicit knowledge of the many-body wave function. Thus, large systems can be tackled efficiently within TDDFT approximations~\cite{Cocchi2014}.
 
Before moving on to discuss an application and thus gain further insights, we emphasize that spin-orbit coupling and magnetic fields are not included in our considerations. We work under the assumption  that the ground state is a spin singlet, thus, both expressions in Eq.~\eqref{CrossSectionEf} and Eq.~\eqref{CrossSecLinEf} allow transitions only within the manifold of \textit{singlet} excited states. Studying the absorption of excited states with different spin multiplicity is important, for example, to account for inter-system crossing~\cite{marian2012spin} which can occur  in optical limiting processes. Formally, this would require to use spin-dependent impulses \cite{Oliveira2008, magnons}.

\section{Analysis of Nonlinear Absorption in a 1D Model System}\label{SecEfAppl}

Here, we  scrutinize the information that can be retrieved by the  ``real-time impulsive method'' on the cross section of an interacting system beyond the linear regime.  Our choice to work at the level of a simple model system, instead of a real molecule, allows us to avoid from the outset the challenge of the typical approximations of state-of-the-art TDDFT. Below, we also compare results from the perturbative expressions derived in the previous section with the computed non-perturbative solution obtained by directly evolving the many-body sate in time.

The considered system consists of two interacting electrons confined in a one-dimensional segment by a potential well of infinite depth (hereafter 1DW). The unperturbed Hamiltonian of the 1DW is
\begin{equation}\label{Eq1DWHam}
\Hzero = \sum_{i=1}^2 \left[ -\frac{1}{2}\frac{\partial^2}{\partial x_i^2}+ v_{ext}(x_i)\right]+\frac{1}{\sqrt{1+(x_1-x_2)^2}} \ ,
\end{equation}
with the external potential
\begin{equation}\label{v01Dwell}
v_{ext}(x_i) = \begin{cases} 0 & -L/2 < x_i < L/2 \\ \infty & \mbox{otherwise} \end{cases} \ .
\end{equation}
The second term in Eq.~\eqref{Eq1DWHam} is the Coulomb interaction between the two electrons, which is softened to avoid the singularity at $x_1=x_2$\cite{Su1991}. For the numerical simulation, we have employed the Octopus code~\cite{Octopus1,numdetails}. $\Hzero$ is symmetric under particle interchange $x_1 \leftrightarrow x_2$. Hence, we can choose the spatial component of the wavefunction to be either symmetric or antisymmetric with respect to the exchange of the spatial coordinates. The eigenstates belong to the irreducible representation of either singlet or triplet spin multiplicity. In addition, $\Hzero$ has also spatial inversion symmetry. Consequently, the orbital part of the wavefunctions must be either even or odd under inversion of the coordinates. 


\begin{figure}[ht]
\begin{center}
\includegraphics[scale=1.00]{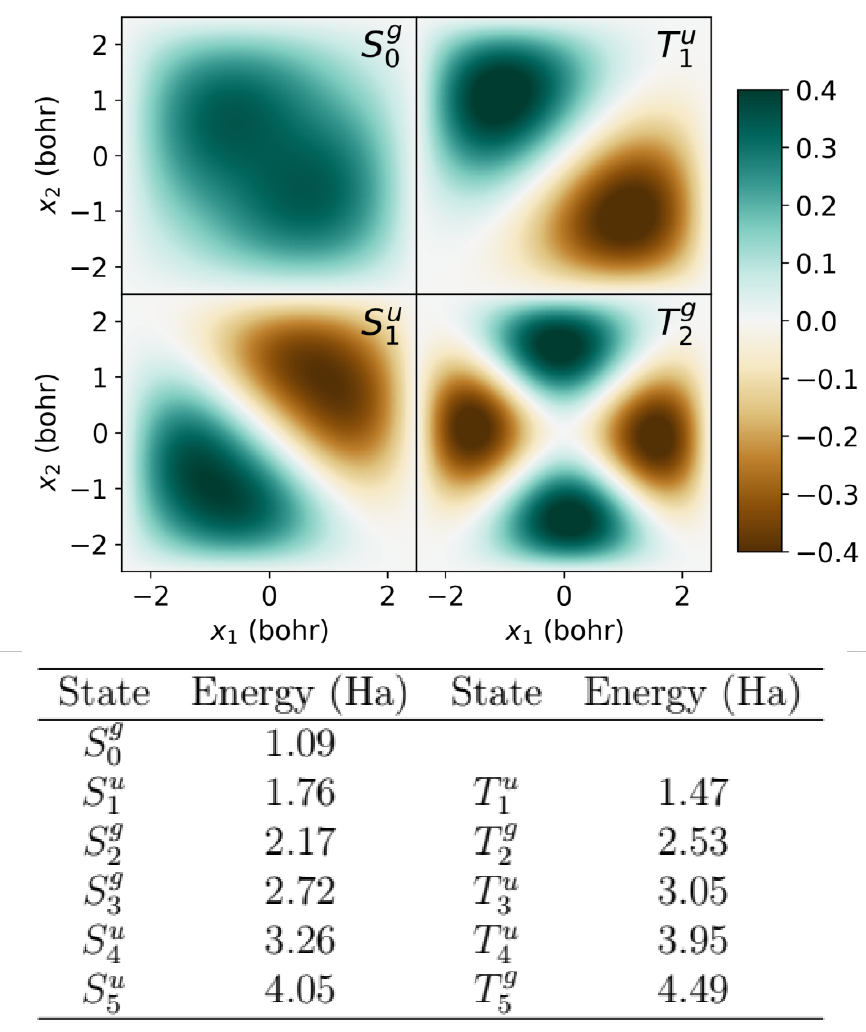}
\caption{Top panel: Eigenstates depicted in the configuration space $x_1-x_2$ (left) and eigenvalues (right) of the unperturbed system of two electrons in a 1D potential well of infinite depth (Eq.~\eqref{Eq1DWHam}). The states are labeled as ${\cal S}_i^{g/u}$, where ${\cal S}$ indicates the spin state (singlet $S$ or triplet $T$), $i$ the order in energy within the spin channel,  and $g/u$ the parity with respect to inversion of the coordinates. Colorbar units are bohr$^{-1}$. Bottom panel: Eigenvalues of the first five  eigenvectors of the unperturbed Hamiltonian in Eq.~\eqref{Eq1DWHam}.}
\label{FigStates}
\end{center}
\end{figure}

The time-dependent polarization is calculated as $d(t) = \mel{\Psi(t)}{\hat{d}}{\Psi(t)}$ and its Fourier transform $\tilde{d}(\omega)$ enters the absorption cross section
\begin{equation}\label{DefCrSec1D}
\sigma(\omega) = \frac{4\pi\omega}{c}\frac{\Im[\tilde{d}(\omega)]}{\kappa} \ .
\end{equation}
The eigenstates and eigenfunctions of this system are shown in Figure~\ref{FigStates}. The singlet and triplet states are labeled as $S_i$ and $T_i$, respectively, where the subscript $i$ labels the ground state ($i=0$) as well as the excited states ($i>0$). The superscripts $g$ (\textit{gerade}) and $u$ (\textit{ungerade}) indicate the parity of the wavefunction. Due to the parity selection rules, dipole transitions are only allowed between $g$ and $u$ states. In addition, the spin selection rule $\Delta S = 0$ holds, as the perturbed Hamiltonian is spin-independent. Thus, the only allowed transitions are $S_i^g \rightarrow S_j^u$ and $S_i^u \rightarrow S_j^g$ in both the linear and nonlinear regimes.


\begin{figure}[ht]
\begin{center}
\includegraphics[scale=1.00]{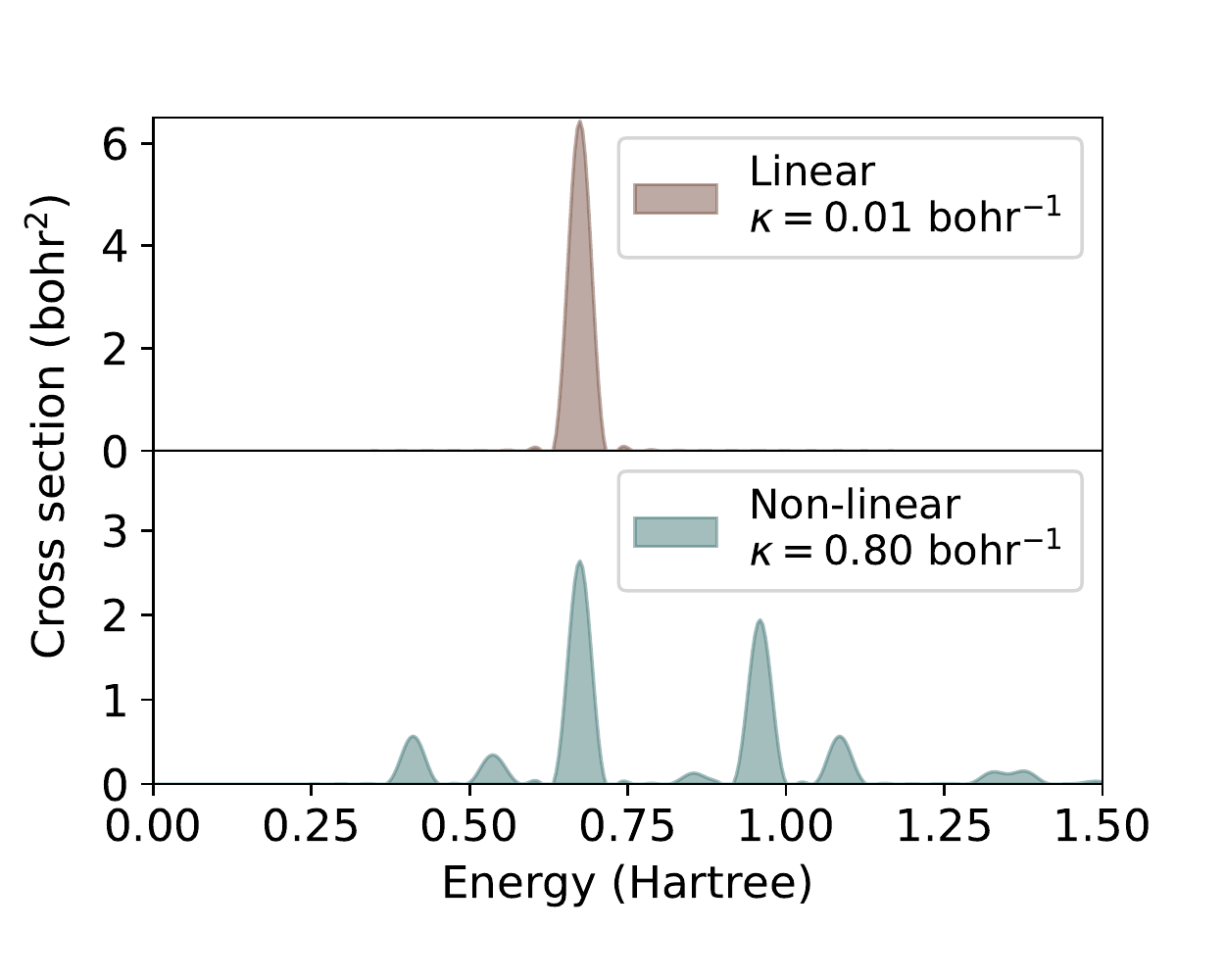}
\caption{Comparison between the linear and nonlinear absorption spectra of a 1D square potential well containing two electrons (1DW). The linear absorption spectrum (upper panel) is obtained by applying an impulsive electric field with strength $\kappa=0.01$ bohr$^{-1}$ (weak-field regime). The nonlinear absorption spectrum (lower panel) is obtained by applying an electric field with strength $\kappa=0.80$ bohr$^{-1}$ (strong field regime).}
\label{FigFirstSpectrum}
\end{center}
\end{figure}

The linear and nonlinear absorption spectra of the 1DW obtained by applying an electric field impulse with $\kappa=0.01$~bohr$^{-1}$ and $0.80$~bohr$^{-1}$, respectively, are shown in Fig.~\ref{FigFirstSpectrum}. Given the well length and the spacing of the ground-state eigenvalues, the field corresponding to $\kappa \le 0.02$ bohr$^{-1}$ can be considered weak. The resulting cross sections show an intrinsic broadening of $0.04$ Ha due to the finite duration of the time propagation.

In the lower panel of Fig. \ref{FigFirstSpectrum} the linear spectrum shows a single maximum at 0.67~Ha, corresponding to the $S_0^g \rightarrow S_1^u$ transition. In contrast, the nonlinear absorption cross section features several peaks spread over the range (0.4--1.4) Ha, namely at lower and higher energies with respect to the excitation in the linear regime. In the nonlinear regime, the maximum corresponding to the $S_0^g \rightarrow S_1^u$ transition at 0.67~Ha is not suppressed but its spectral weight is approximately halved.


\begin{figure}[ht]
\begin{center}
\includegraphics[scale=0.90]{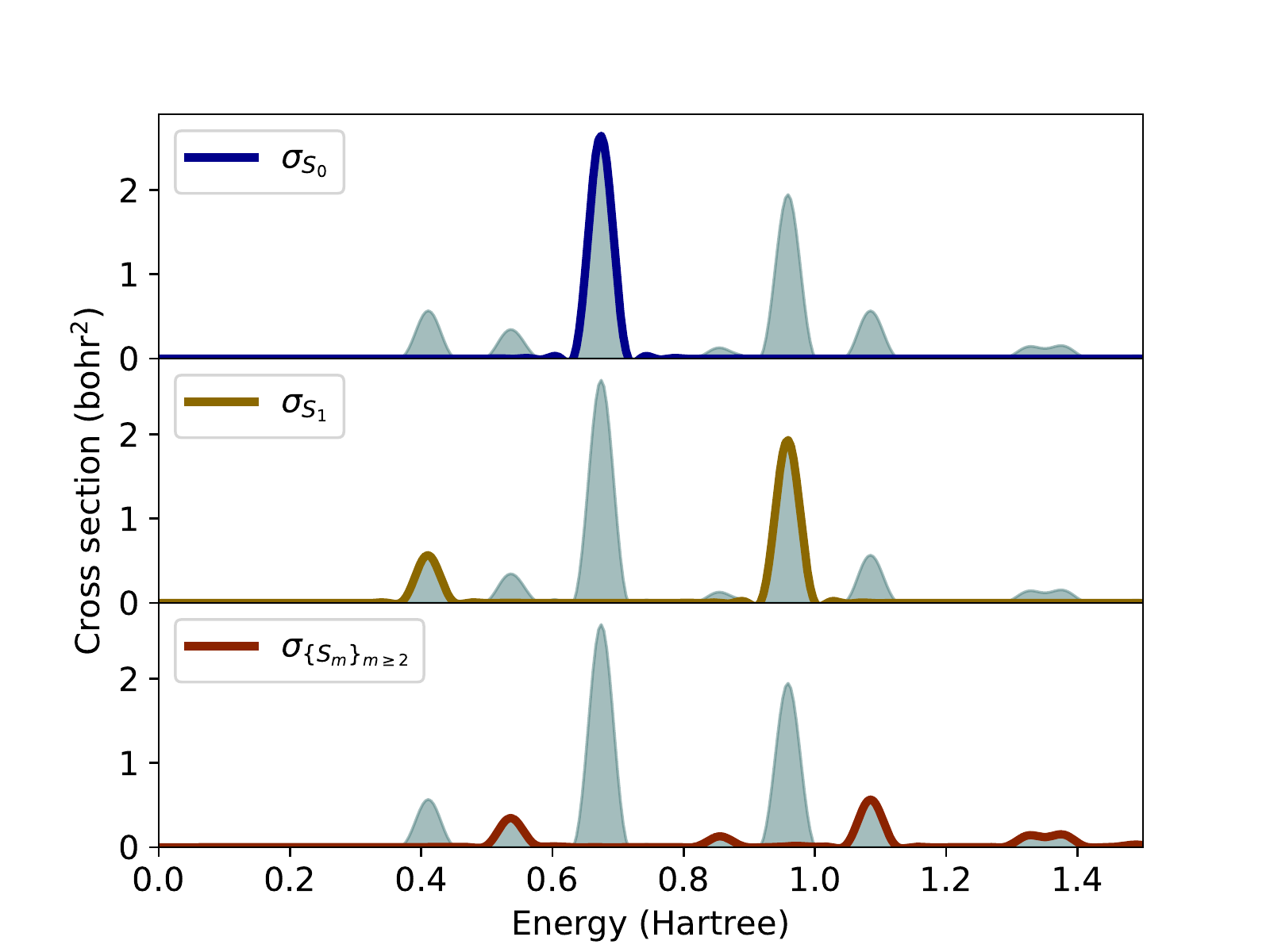}
\caption{Nonlinear absorption cross section of the 1DW subject to a strong electric field impulse with $\kappa=0.80$ bohr$^{-1}$. The cross section is split into ground-state absorption (blue curve), first excited-state absorption (yellow curve) and absorption from higher excited states (red curve). }
\label{FigSpectrumK05}
\end{center}
\end{figure}

To analyze the transitions involved in the nonlinear cross section, in  Fig.~\ref{FigSpectrumK05} we consider separately the contributions of three components, namely $\sigma_{S_0}(\omega)$, $\sigma_{S_1}(\omega)$, and $\sigma_{\lbrace S_m \rbrace}(\omega)$, with $m \ge 2$. 
The first and second components account for the absorption from the ground state and from the first excited state, while the third one includes the contributions from all higher excited states. These components are calculated from Eq.~\eqref{CrossSectionEf} evaluating the sums up to the first $100$ eigenstates of $\Hzero$. Convergence is ensured by the sum rule in Eq.~\eqref{EfSumRule}. Dirac deltas in Eq.~\eqref{CrossSectionEf} are broadened in order to match the peak width of the cross sections obtained from the solution of the time-dependent Schr\"odinger equation.

The component $\sigma_{S_0}(\omega)$ (top panel of Fig.~\ref{FigSpectrumK05}) includes the same contribution as the linear cross section, $S_0^g \rightarrow S_1^u$ (see top panel of Fig.~\ref{FigFirstSpectrum}). Therefore, the additional peaks in the lower panel of Fig.~\ref{FigFirstSpectrum} result from ESA.
In particular, the two peaks at 0.41~Ha and 0.96~Ha (see middle panel of Fig.~\ref{FigSpectrumK05}) are due to the absorption from the first excited state ($S_1^u$) and involve the transitions to \textit{gerade} excited states $S_1^u \rightarrow S_2^g$ and $S_1^u \rightarrow S_3^g$, respectively. Simply based on symmetry, the excited states $S_2^g$ and $S_3^g$ cannot be reached from the ground state $S_0^g$. By inspecting the bottom panel of Fig.~\ref{FigSpectrumK05}, we notice that the higher-order contributions to the absorption consist of a number of weak peaks below and above the maximum at 0.67~Ha. For example, the maximum at about 1.1~Ha corresponds to the transition $S_2^g \rightarrow S_4^u$.

%
\begin{figure}[ht]
\begin{center}
\includegraphics[scale=0.50]{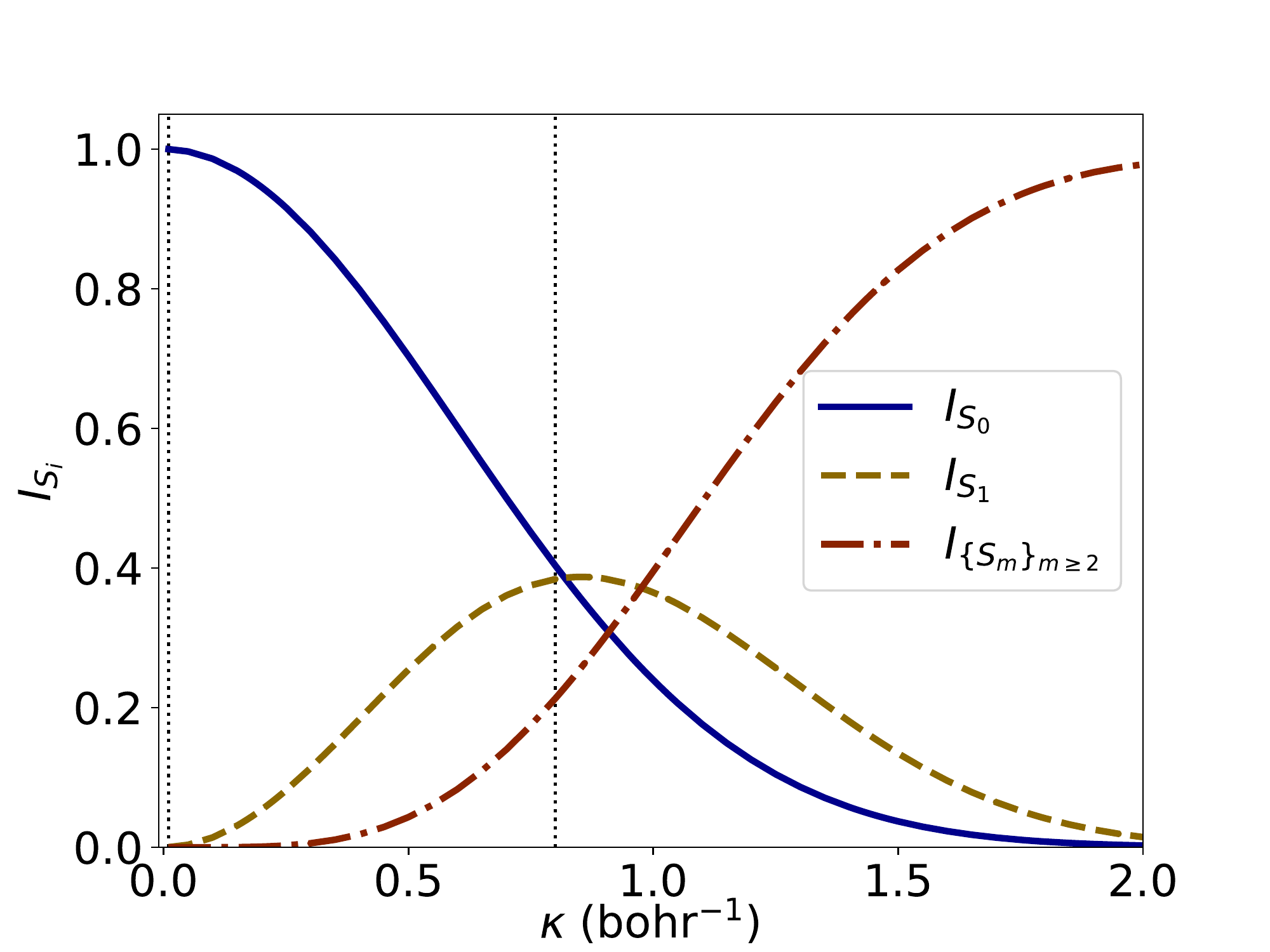}
\caption{Normalized weights [see Eq.~\eqref{RecWeigthEf}] of the three components of the absorption cross section given in Figure~\ref{FigSpectrumK05} as a function of the strength of the electric field impulse $\kappa$. The dashed vertical bars correspond to the values of $\kappa$ used in Figure~\ref{FigFirstSpectrum}.}
\label{FigMelKick}
\end{center}
\end{figure}

To gain further understanding on the information provided by the non-perturbative approach presented so far, it is interesting to inspect the variation of the relative weights of GSA and ESA as a function of the field strength. 
These contributions are quantified by
\begin{equation}\label{RecWeigthEf}
I_{S_i}=\frac{c}{4\pi^2}\int\limits_{0}^{+\infty} d\omega \ \sigma_{S_i}(\omega) \ .
\end{equation}
The values of $I_{S_0}$, $I_{S_1}$, and $I_{S_m}$, with $m\ge 2$, are shown in Fig.~\ref{FigMelKick} as a function of $\kappa$. 
Note that each contribution refers to a specific subset of the absorption but they all include contributions {\em at all perturbation orders}.
The solid curve in Fig.~\ref{FigMelKick} represents the weight of the ground-state cross section $I_{S_0}$. For $\kappa \approx 0$, we have that $I_{S_0} \approx 1$, which confirms that there is only GSA in the linear regime, as expected. 
At increasing values of $\kappa$, $I_{S_0}$ decreases monotonically: ESA becomes relevant as the response of the system deviates from linearity.
For $\kappa > 1.44$~bohr$^{-1}$, we have that $I_{S_0} < 0.05$, meaning that GSA becomes negligible above this threshold. 
The dashed curve in Fig.~\ref{FigMelKick} represents the weight of the first-excited-state cross section $I_{S_1}$. 
It vanishes for small $\kappa$, reaches its maximum at $\kappa = 0.85$~Ha, and decreases monotonically for higher values of $\kappa$. 
The dashed-dotted curve in Fig.~\ref{FigMelKick} accounts for the weight of the higher-order absorption cross section $I_{\lbrace S_m\rbrace}$, where $m \ge 2$. 
Also $I_{\lbrace S_m\rbrace}$ does not contribute at small $\kappa$.
It increases monotonically starting from $\kappa \approx 0.25$ bohr$^{-1}$ and reaches its saturation value, $I_{\lbrace S_m\rbrace} = 1$, for $\kappa$ approaching 2~bohr$^{-1}$.
In the strong field limit ($\kappa > 1.75$~bohr$^{-1}$), the only non-negligible component of the cross section is $I_{\lbrace S_m \rbrace}$. 

%
\begin{figure}[ht!]
\begin{center}
\includegraphics[scale=0.46]{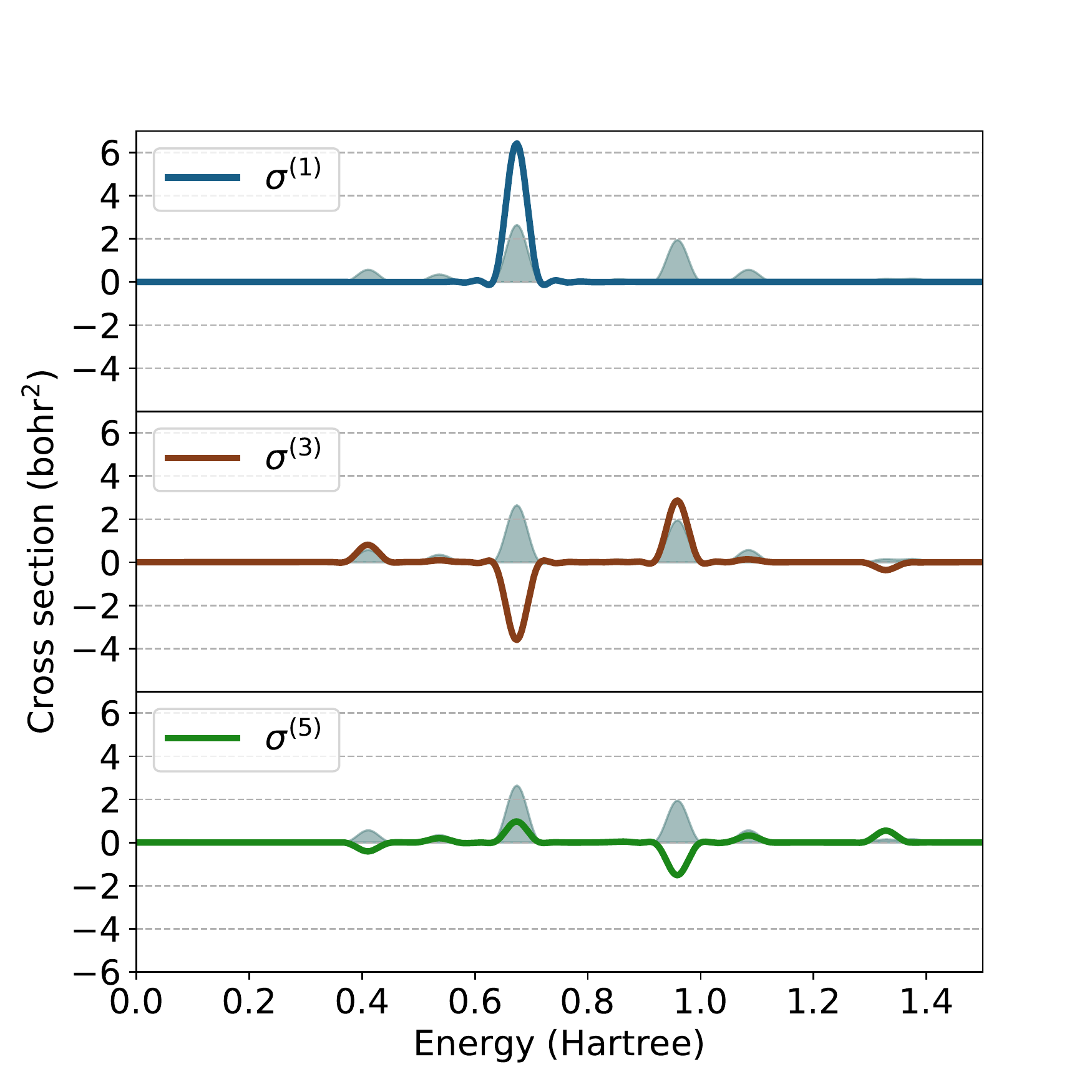}
\includegraphics[scale=0.46]{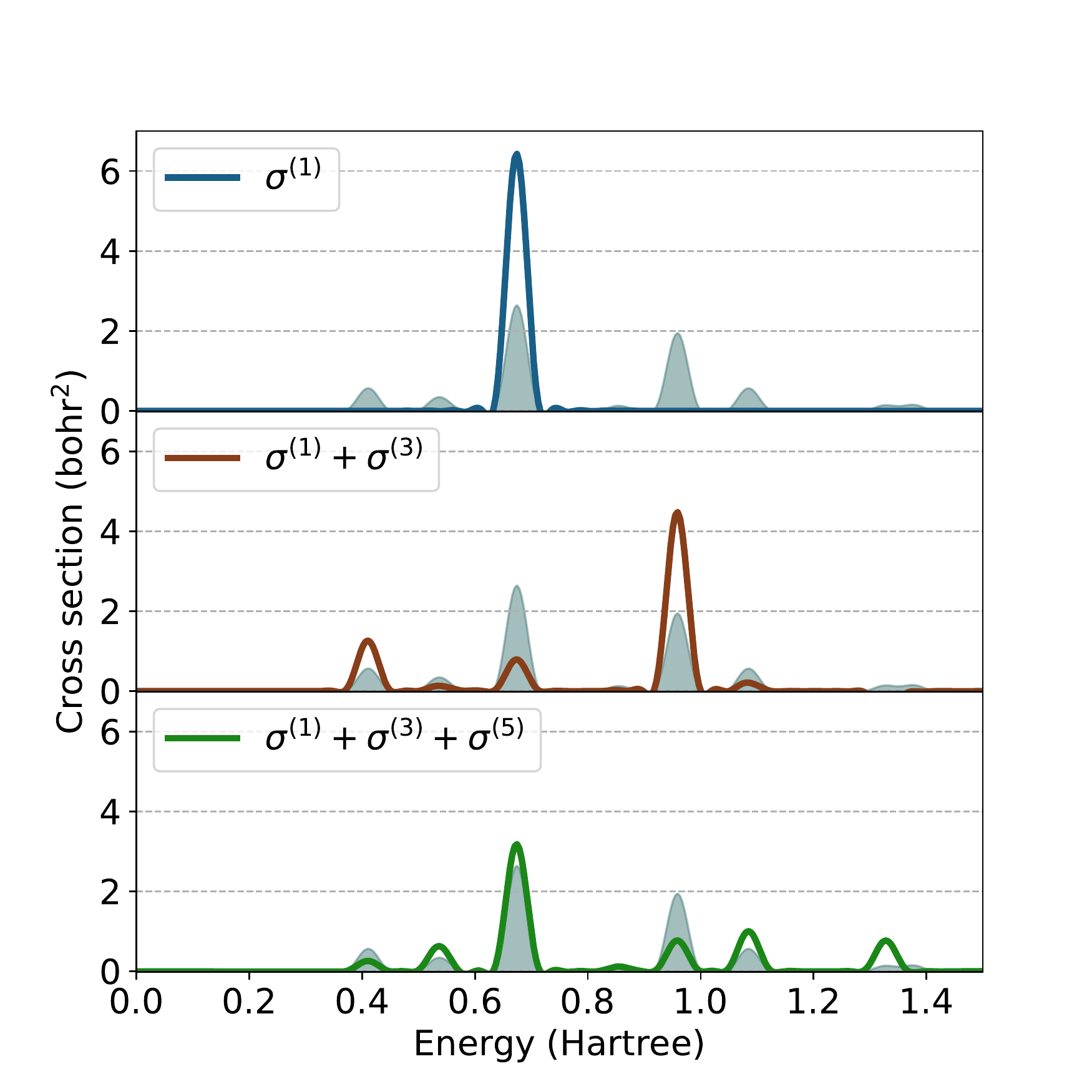}
\caption{Left panel: perturbative components $\sigma^{(1)}$, $\sigma^{(3)}$ and $\sigma^{(5)}$ (Eqs.~\eqref{CrossSecLinEf}, \eqref{sigma3impulse} and \eqref{sigma5imp} respectively) of the nonlinear absorption cross section the 1DW subject to the electric field impulse $\kappa = 0.80$ bohr$^{-1}$. For reference, the total cross section (Eq. \ref{CrossSectionEf}), also shown in the bottom panel of Fig.~\ref{FigFirstSpectrum}, is plotted as shaded area in the background of each panel. Right panel: summation of the first five non-zero terms in the perturbation series compared to the full non-perturbative result (shaded grey area).}
\label{PertSpectra}
\end{center}
\end{figure}

We complete our analysis by considering the perturbative expansion of the nonlinear absorption cross section, as discussed in Section~\ref{SecEf}. 
The perturbative terms $\sigma^{(1)}(\omega)$, $\sigma^{(3)}(\omega)$, and $\sigma^{(5)}(\omega)$ are shown in on the left panel of Fig.~\ref{PertSpectra}. They are computed from Eq.~\eqref{CrossSecLinEf}, Eq.~\eqref{sigma3impulse}, and Eq.~\eqref{sigma5imp}. The cross section $\sigma(\omega)$ calculated with the impulse response method for $\kappa = 0.80$ bohr$^{-1}$ is also shown for reference. For comparison, on the right panel of Fig.~\ref{PertSpectra} we show the perturbative contribution \textit{summed up} to the indicated order.

The first-order cross section $\sigma^{(1)}(\omega)$ assumes only positive values and contributes only to the maximum at 0.67 Ha.
The first nonlinear non-vanishing component of the cross section, which thus accounts for ESA processes, is the third-order one, $\sigma^{(3)}$, which assumes both positive and negative values. A pronounced peak with negative strength is found at 0.67 Ha and corresponds to the third-order correction of the ground-state transition $S_0^g \rightarrow S_1^u$. This term cancels out almost completely the contribution from $\sigma^{(1)}$ at the same energy. Maxima with positive intensities appear at about 0.4~Ha and 1.0~Ha.

However, the contribution of $\sigma^{(3)}$ alone is not sufficient to describe the nonlinear excitations in the 1DW -- 
see Fig.~\ref{PertSpectra}, left panel, middle graph. For this purpose, it is necessary to include at least also the contributions from the fifth-order cross section, $\sigma^{(5)}$. This term has maxima and minima at the same energy as those of $\sigma^{(3)}$ but with intensities of opposite sign. This is a general feature of perturbation theory. In particular, the peak at 0.67~Ha is positive and overlaps almost perfectly with the one in the impulse cross section shown in the background. Minima are found at approximately 0.4~Ha and 1.0~Ha, at the same energies where $\sigma^{(3)}$ exhibits maxima. The absolute values of the corresponding intensities are very similar, suggesting that these contributions should cancel out.

Before moving to the general conclusions, we stress the fact that summing up the perturbative contributions of $\sigma$ up to the fifth order is still not sufficient to match the non-perturbative result. Even assuming that the perturbative series is within its convergence radius (which is not trivially granted), Fig.~\ref{PertSpectra} shows that for large values of $\kappa$ the terms of the series tend to have alternating sign contributions, which explains the observed difficulties in the convergence.

\section{Conclusions}

Our analysis demonstrates that the time-evolution of many-electron systems induced by an electrical field in the instantaneous limit is an effective tool for investigating computationally fluence-dependent nonlinear optical properties. It works well also for those cases in which the convergence of the perturbative expansions of the cross sections is challenging. Looking ahead, nonlinearities arising from nuclear motion can be straightforwardly included via hybridization with a scheme for the dynamics of the nuclei. 

We also stress that the considered method cannot access the dependence of the spectrum on the pulse shape. Therefore, it is not useful, as it is, to study processes such as two-photon absorption or ultrafast transients. It is not suitable either to investigate those cases in which hysteresis loops or other memory-dependent phenomena are important.

To assess the capabilities of the method itself independently of other approximations -- such as those intrinsically entailed, for example, in (TD)DFT -- we have studied a 1D model system to numerically verify our findings using a finite-differences propagation scheme in real space and real time.
The approach examined in this work provides us with vital information about the whole spectrum at once. This is particularly relevant for investigating nonlinear systems, for which the total absorption cannot be decomposed into the sum of the individual absorption of monochromatic radiation for each frequency. Nonlinearities arising from nuclear motion can be straightforwardly included via hybridization with a scheme for the dynamics of the nuclei. 

We conclude that the impulse response as computed in real time can profitably be employed to study optical nonlinearities. Specifically, we have shown that the impulsive method provides relevant information about the steady-state absorption of time-invariant systems in which the non linearity manifests itself as a mere dependence of the spectrum on the field strength. 
For cases in which the interest in optical nonlinearities is not restricted to wave mixing at a predefined order, the computation of the nonlinear impulsive response function from real-time propagation can outperform the ordinary approach based on perturbation theory to investigate phenomena driven by  excited state absorption such as reverse saturable absorption, optical switching, and optical limiting.

\begin{acknowledgement}
A.G. acknowledges financial support from the German Academic Exchange Service (DAAD) grant n. 57440917 and from HPC Europa 3 grant n. HPC17AS2HO. 
C.C. appreciates funding from the German Research Foundation (DFG), Project number 182087777 -
SFB 951.
Computational resources provided by the North-German Supercomputing Alliance (HLRN), project bep00060, and by the High Performance Computing Center Stuttgart (HLRS). 
\end{acknowledgement}

\begin{suppinfo}
In Sec. \ref{SecCrossSectAppx} the general expression for the nonlinear absorption cross section is derived. Ref.~\citenum{Jackson} is cited therein. In Sec. \ref{SecOptGapAppx} the spectral resolution of the cross section is derived.  Sec.~\ref{Secsigma5Appx} $\sigma^{(5)}$ is split in GSA and ESA components.
\end{suppinfo}

\bibliography{Bibliography}

\providecommand{\latin}[1]{#1}
\makeatletter
\providecommand{\doi}
  {\begingroup\let\do\@makeother\dospecials
  \catcode`\{=1 \catcode`\}=2 \doi@aux}
\providecommand{\doi@aux}[1]{\endgroup\texttt{#1}}
\makeatother
\providecommand*\mcitethebibliography{\thebibliography}
\csname @ifundefined\endcsname{endmcitethebibliography}
  {\let\endmcitethebibliography\endthebibliography}{}
\begin{mcitethebibliography}{96}
\providecommand*\natexlab[1]{#1}
\providecommand*\mciteSetBstSublistMode[1]{}
\providecommand*\mciteSetBstMaxWidthForm[2]{}
\providecommand*\mciteBstWouldAddEndPuncttrue
  {\def\EndOfBibitem{\unskip.}}
\providecommand*\mciteBstWouldAddEndPunctfalse
  {\let\EndOfBibitem\relax}
\providecommand*\mciteSetBstMidEndSepPunct[3]{}
\providecommand*\mciteSetBstSublistLabelBeginEnd[3]{}
\providecommand*\EndOfBibitem{}
\mciteSetBstSublistMode{f}
\mciteSetBstMaxWidthForm{subitem}{(\alph{mcitesubitemcount})}
\mciteSetBstSublistLabelBeginEnd
  {\mcitemaxwidthsubitemform\space}
  {\relax}
  {\relax}

\bibitem[Franken \latin{et~al.}(1961)Franken, Hill, Peters, and
  Weinreich]{Franken1961}
Franken,~P.~A.; Hill,~A.~E.; Peters,~C.~W.; Weinreich,~G. Generation of Optical
  Harmonics. \emph{Phys. Rev. Lett.} \textbf{1961}, \emph{7}, 118--119\relax
\mciteBstWouldAddEndPuncttrue
\mciteSetBstMidEndSepPunct{\mcitedefaultmidpunct}
{\mcitedefaultendpunct}{\mcitedefaultseppunct}\relax
\EndOfBibitem
\bibitem[Damm \latin{et~al.}(1985)Damm, Kaschke, Noack, and Wilhelmi]{Damm1985}
Damm,~T.; Kaschke,~M.; Noack,~F.; Wilhelmi,~B. Compression of picosecond pulses
  from a solid-state laser using self-phase modulation in graded-index fibers.
  \emph{Opt. Lett.} \textbf{1985}, \emph{10}, 176--178\relax
\mciteBstWouldAddEndPuncttrue
\mciteSetBstMidEndSepPunct{\mcitedefaultmidpunct}
{\mcitedefaultendpunct}{\mcitedefaultseppunct}\relax
\EndOfBibitem
\bibitem[Perry and Mourou(1994)Perry, and Mourou]{Perry1994}
Perry,~M.~D.; Mourou,~G. Terawatt to Petawatt Subpicosecond Lasers.
  \emph{Science} \textbf{1994}, \emph{264}, 917--924\relax
\mciteBstWouldAddEndPuncttrue
\mciteSetBstMidEndSepPunct{\mcitedefaultmidpunct}
{\mcitedefaultendpunct}{\mcitedefaultseppunct}\relax
\EndOfBibitem
\bibitem[Boyd(2008)]{Boyd2008}
Boyd,~R.~W. \emph{Nonlinear Optics, Third Edition}, 3rd ed.; Academic Press,
  Inc.: USA, 2008\relax
\mciteBstWouldAddEndPuncttrue
\mciteSetBstMidEndSepPunct{\mcitedefaultmidpunct}
{\mcitedefaultendpunct}{\mcitedefaultseppunct}\relax
\EndOfBibitem
\bibitem[Tretiak and Chernyak(2003)Tretiak, and Chernyak]{Chernyak_2003}
Tretiak,~S.; Chernyak,~V. Resonant nonlinear polarizabilities in the
  time-dependent density functional theory. \emph{The Journal of Chemical
  Physics} \textbf{2003}, \emph{119}, 8809--8823\relax
\mciteBstWouldAddEndPuncttrue
\mciteSetBstMidEndSepPunct{\mcitedefaultmidpunct}
{\mcitedefaultendpunct}{\mcitedefaultseppunct}\relax
\EndOfBibitem
\bibitem[Grimberg \latin{et~al.}(2002)Grimberg, Lozovoy, Dantus, and
  Mukamel]{Mukamel_2001}
Grimberg,~B.~I.; Lozovoy,~V.~V.; Dantus,~M.; Mukamel,~S. Ultrafast Nonlinear
  Spectroscopic Techniques in the Gas Phase and Their Density Matrix
  Representation. \emph{The Journal of Physical Chemistry A} \textbf{2002},
  \emph{106}, 697--718\relax
\mciteBstWouldAddEndPuncttrue
\mciteSetBstMidEndSepPunct{\mcitedefaultmidpunct}
{\mcitedefaultendpunct}{\mcitedefaultseppunct}\relax
\EndOfBibitem
\bibitem[Tunell \latin{et~al.}(2003)Tunell, Rinkevicius, Vahtras, Sałek,
  Helgaker, and Ågren]{Agren_2003}
Tunell,~I.; Rinkevicius,~Z.; Vahtras,~O.; Sałek,~P.; Helgaker,~T.; Ågren,~H.
  Density functional theory of nonlinear triplet response properties with
  applications to phosphorescence. \emph{The Journal of Chemical Physics}
  \textbf{2003}, \emph{119}, 11024--11034\relax
\mciteBstWouldAddEndPuncttrue
\mciteSetBstMidEndSepPunct{\mcitedefaultmidpunct}
{\mcitedefaultendpunct}{\mcitedefaultseppunct}\relax
\EndOfBibitem
\bibitem[Jansik \latin{et~al.}(2005)Jansik, Sałek, Jonsson, Vahtras, and
  Ågren]{Agren_2005}
Jansik,~B.; Sałek,~P.; Jonsson,~D.; Vahtras,~O.; Ågren,~H. Cubic response
  functions in time-dependent density functional theory. \emph{The Journal of
  Chemical Physics} \textbf{2005}, \emph{122}, 054107\relax
\mciteBstWouldAddEndPuncttrue
\mciteSetBstMidEndSepPunct{\mcitedefaultmidpunct}
{\mcitedefaultendpunct}{\mcitedefaultseppunct}\relax
\EndOfBibitem
\bibitem[van Gisbergen \latin{et~al.}(1998)van Gisbergen, Snijders, and
  Baerends]{Gisbergen_1998}
van Gisbergen,~S. J.~A.; Snijders,~J.~G.; Baerends,~E.~J. Calculating
  frequency-dependent hyperpolarizabilities using time-dependent density
  functional theory. \emph{The Journal of Chemical Physics} \textbf{1998},
  \emph{109}, 10644--10656\relax
\mciteBstWouldAddEndPuncttrue
\mciteSetBstMidEndSepPunct{\mcitedefaultmidpunct}
{\mcitedefaultendpunct}{\mcitedefaultseppunct}\relax
\EndOfBibitem
\bibitem[de~Wergifosse and Grimme(2018)de~Wergifosse, and Grimme]{Grimme_2018}
de~Wergifosse,~M.; Grimme,~S. Nonlinear-response properties in a simplified
  time-dependent density functional theory (sTD-DFT) framework: Evaluation of
  the first hyperpolarizability. \emph{The Journal of Chemical Physics}
  \textbf{2018}, \emph{149}, 024108\relax
\mciteBstWouldAddEndPuncttrue
\mciteSetBstMidEndSepPunct{\mcitedefaultmidpunct}
{\mcitedefaultendpunct}{\mcitedefaultseppunct}\relax
\EndOfBibitem
\bibitem[Hait~Heinze \latin{et~al.}(2002)Hait~Heinze, Della~Sala, and
  Görling]{Haitheinze_2002}
Hait~Heinze,~H.; Della~Sala,~F.; Görling,~A. Efficient methods to calculate
  dynamic hyperpolarizability tensors by time-dependent density-functional
  theory. \emph{The Journal of Chemical Physics} \textbf{2002}, \emph{116},
  9624--9640\relax
\mciteBstWouldAddEndPuncttrue
\mciteSetBstMidEndSepPunct{\mcitedefaultmidpunct}
{\mcitedefaultendpunct}{\mcitedefaultseppunct}\relax
\EndOfBibitem
\bibitem[Henriksson \latin{et~al.}(2008)Henriksson, Saue, and
  Norman]{Henriksson_2008}
Henriksson,~J.; Saue,~T.; Norman,~P. Quadratic response functions in the
  relativistic four-component Kohn-Sham approximation. \emph{The Journal of
  Chemical Physics} \textbf{2008}, \emph{128}, 024105\relax
\mciteBstWouldAddEndPuncttrue
\mciteSetBstMidEndSepPunct{\mcitedefaultmidpunct}
{\mcitedefaultendpunct}{\mcitedefaultseppunct}\relax
\EndOfBibitem
\bibitem[Senatore and Subbaswamy(1987)Senatore, and Subbaswamy]{Senatore_1986}
Senatore,~G.; Subbaswamy,~K.~R. Nonlinear response of closed-shell atoms in the
  density-functional formalism. \emph{Phys. Rev. A} \textbf{1987}, \emph{35},
  2440--2447\relax
\mciteBstWouldAddEndPuncttrue
\mciteSetBstMidEndSepPunct{\mcitedefaultmidpunct}
{\mcitedefaultendpunct}{\mcitedefaultseppunct}\relax
\EndOfBibitem
\bibitem[Iwata \latin{et~al.}(2001)Iwata, Yabana, and Bertsch]{Yabana_2001}
Iwata,~J.-I.; Yabana,~K.; Bertsch,~G.~F. Real-space computation of dynamic
  hyperpolarizabilities. \emph{The Journal of Chemical Physics} \textbf{2001},
  \emph{115}, 8773--8783\relax
\mciteBstWouldAddEndPuncttrue
\mciteSetBstMidEndSepPunct{\mcitedefaultmidpunct}
{\mcitedefaultendpunct}{\mcitedefaultseppunct}\relax
\EndOfBibitem
\bibitem[Ye and Autschbach(2006)Ye, and Autschbach]{Ye_2006}
Ye,~A.; Autschbach,~J. Study of static and dynamic first hyperpolarizabilities
  using time-dependent density functional quadratic response theory with local
  contribution and natural bond orbital analysis. \emph{The Journal of Chemical
  Physics} \textbf{2006}, \emph{125}, 234101\relax
\mciteBstWouldAddEndPuncttrue
\mciteSetBstMidEndSepPunct{\mcitedefaultmidpunct}
{\mcitedefaultendpunct}{\mcitedefaultseppunct}\relax
\EndOfBibitem
\bibitem[Kanis \latin{et~al.}(1992)Kanis, Ratner, and Marks]{Kanis_1992}
Kanis,~D.~R.; Ratner,~M.~A.; Marks,~T.~J. Calculation and electronic
  description of quadratic hyperpolarizabilities. Toward a molecular
  understanding of NLO responses in organotransition metal chromophores.
  \emph{Journal of the American Chemical Society} \textbf{1992}, \emph{114},
  10338--10357\relax
\mciteBstWouldAddEndPuncttrue
\mciteSetBstMidEndSepPunct{\mcitedefaultmidpunct}
{\mcitedefaultendpunct}{\mcitedefaultseppunct}\relax
\EndOfBibitem
\bibitem[Quinet \latin{et~al.}(2001)Quinet, Champagne, and
  Kirtman]{Quinet_2001}
Quinet,~O.; Champagne,~B.; Kirtman,~B. Analytical TDHF second derivatives of
  dynamic electronic polarizability with respect to nuclear coordinates.
  Application to the dynamic ZPVA correction. \emph{Journal of Computational
  Chemistry} \textbf{2001}, \emph{22}, 1920--1932\relax
\mciteBstWouldAddEndPuncttrue
\mciteSetBstMidEndSepPunct{\mcitedefaultmidpunct}
{\mcitedefaultendpunct}{\mcitedefaultseppunct}\relax
\EndOfBibitem
\bibitem[Quinet and Champagne(2001)Quinet, and Champagne]{Quinet_2001_2}
Quinet,~O.; Champagne,~B. Sum-frequency generation first hyperpolarizability
  from time-dependent Hartree–Fock method. \emph{International Journal of
  Quantum Chemistry} \textbf{2001}, \emph{85}, 463--468\relax
\mciteBstWouldAddEndPuncttrue
\mciteSetBstMidEndSepPunct{\mcitedefaultmidpunct}
{\mcitedefaultendpunct}{\mcitedefaultseppunct}\relax
\EndOfBibitem
\bibitem[Quinet and Champagne(2002)Quinet, and Champagne]{Quinet_2002}
Quinet,~O.; Champagne,~B. Analytical time-dependent Hartree–Fock schemes for
  the evaluation of the hyper-Raman intensities. \emph{The Journal of Chemical
  Physics} \textbf{2002}, \emph{117}, 2481--2488\relax
\mciteBstWouldAddEndPuncttrue
\mciteSetBstMidEndSepPunct{\mcitedefaultmidpunct}
{\mcitedefaultendpunct}{\mcitedefaultseppunct}\relax
\EndOfBibitem
\bibitem[Jonsson \latin{et~al.}(1996)Jonsson, Norman, and Ågren]{Agren_1996}
Jonsson,~D.; Norman,~P.; Ågren,~H. Cubic response functions in the
  multiconfiguration self‐consistent field approximation. \emph{The Journal
  of Chemical Physics} \textbf{1996}, \emph{105}, 6401--6419\relax
\mciteBstWouldAddEndPuncttrue
\mciteSetBstMidEndSepPunct{\mcitedefaultmidpunct}
{\mcitedefaultendpunct}{\mcitedefaultseppunct}\relax
\EndOfBibitem
\bibitem[Berman and Mukamel(2003)Berman, and Mukamel]{Mukamel_2003}
Berman,~O.; Mukamel,~S. Quasiparticle density-matrix representation of
  nonlinear time-dependent density-functional response functions. \emph{Phys.
  Rev. A} \textbf{2003}, \emph{67}, 042503\relax
\mciteBstWouldAddEndPuncttrue
\mciteSetBstMidEndSepPunct{\mcitedefaultmidpunct}
{\mcitedefaultendpunct}{\mcitedefaultseppunct}\relax
\EndOfBibitem
\bibitem[Hettema \latin{et~al.}(1992)Hettema, Jensen, Jo/rgensen, and
  Olsen]{Olsen_1992}
Hettema,~H.; Jensen,~H. J.~A.; Jo/rgensen,~P.; Olsen,~J. Quadratic response
  functions for a multiconfigurational self‐consistent field wave function.
  \emph{The Journal of Chemical Physics} \textbf{1992}, \emph{97},
  1174--1190\relax
\mciteBstWouldAddEndPuncttrue
\mciteSetBstMidEndSepPunct{\mcitedefaultmidpunct}
{\mcitedefaultendpunct}{\mcitedefaultseppunct}\relax
\EndOfBibitem
\bibitem[Ye \latin{et~al.}(2007)Ye, Patchkovskii, and Autschbach]{Ye_2007}
Ye,~A.; Patchkovskii,~S.; Autschbach,~J. Static and dynamic second
  hyperpolarizability calculated by time-dependent density functional cubic
  response theory with local contribution and natural bond orbital analysis.
  \emph{The Journal of Chemical Physics} \textbf{2007}, \emph{127},
  074104\relax
\mciteBstWouldAddEndPuncttrue
\mciteSetBstMidEndSepPunct{\mcitedefaultmidpunct}
{\mcitedefaultendpunct}{\mcitedefaultseppunct}\relax
\EndOfBibitem
\bibitem[Sałek \latin{et~al.}(2002)Sałek, Vahtras, Helgaker, and
  Ågren]{Agren_2002}
Sałek,~P.; Vahtras,~O.; Helgaker,~T.; Ågren,~H. Density-functional theory of
  linear and nonlinear time-dependent molecular properties. \emph{The Journal
  of Chemical Physics} \textbf{2002}, \emph{117}, 9630--9645\relax
\mciteBstWouldAddEndPuncttrue
\mciteSetBstMidEndSepPunct{\mcitedefaultmidpunct}
{\mcitedefaultendpunct}{\mcitedefaultseppunct}\relax
\EndOfBibitem
\bibitem[Rinkevicius \latin{et~al.}(2007)Rinkevicius, Jha, Oprea, Vahtras, and
  Ågren]{Agren_2007}
Rinkevicius,~Z.; Jha,~P.~C.; Oprea,~C.~I.; Vahtras,~O.; Ågren,~H.
  Time-dependent density functional theory for nonlinear properties of
  open-shell systems. \emph{The Journal of Chemical Physics} \textbf{2007},
  \emph{127}, 114101\relax
\mciteBstWouldAddEndPuncttrue
\mciteSetBstMidEndSepPunct{\mcitedefaultmidpunct}
{\mcitedefaultendpunct}{\mcitedefaultseppunct}\relax
\EndOfBibitem
\bibitem[Mai(2011)]{Maitra_2011}
Perspectives on double-excitations in TDDFT. \emph{Chemical Physics}
  \textbf{2011}, \emph{391}, 110 -- 119\relax
\mciteBstWouldAddEndPuncttrue
\mciteSetBstMidEndSepPunct{\mcitedefaultmidpunct}
{\mcitedefaultendpunct}{\mcitedefaultseppunct}\relax
\EndOfBibitem
\bibitem[Parker \latin{et~al.}(2018)Parker, Rappoport, and Furche]{Parker_2017}
Parker,~S.~M.; Rappoport,~D.; Furche,~F. Quadratic Response Properties from
  TDDFT: Trials and Tribulations. \emph{Journal of Chemical Theory and
  Computation} \textbf{2018}, \emph{14}, 807--819\relax
\mciteBstWouldAddEndPuncttrue
\mciteSetBstMidEndSepPunct{\mcitedefaultmidpunct}
{\mcitedefaultendpunct}{\mcitedefaultseppunct}\relax
\EndOfBibitem
\bibitem[Andrade \latin{et~al.}(2007)Andrade, Botti, Marques, and
  Rubio]{Andrade_2007}
Andrade,~X.; Botti,~S.; Marques,~M. A.~L.; Rubio,~A. Time-dependent density
  functional theory scheme for efficient calculations of dynamic
  (hyper)polarizabilities. \emph{The Journal of Chemical Physics}
  \textbf{2007}, \emph{126}, 184106\relax
\mciteBstWouldAddEndPuncttrue
\mciteSetBstMidEndSepPunct{\mcitedefaultmidpunct}
{\mcitedefaultendpunct}{\mcitedefaultseppunct}\relax
\EndOfBibitem
\bibitem[Luppi \latin{et~al.}(2010)Luppi, H\"ubener, and V\'eniard]{Luppi_2010}
Luppi,~E.; H\"ubener,~H.; V\'eniard,~V. Ab initio second-order nonlinear optics
  in solids: Second-harmonic generation spectroscopy from time-dependent
  density-functional theory. \emph{Phys. Rev. B} \textbf{2010}, \emph{82},
  235201\relax
\mciteBstWouldAddEndPuncttrue
\mciteSetBstMidEndSepPunct{\mcitedefaultmidpunct}
{\mcitedefaultendpunct}{\mcitedefaultseppunct}\relax
\EndOfBibitem
\bibitem[Hughes and Sipe(1996)Hughes, and Sipe]{Hughes_1996}
Hughes,~J. L.~P.; Sipe,~J.~E. Calculation of second-order optical response in
  semiconductors. \emph{Phys. Rev. B} \textbf{1996}, \emph{53},
  10751--10763\relax
\mciteBstWouldAddEndPuncttrue
\mciteSetBstMidEndSepPunct{\mcitedefaultmidpunct}
{\mcitedefaultendpunct}{\mcitedefaultseppunct}\relax
\EndOfBibitem
\bibitem[Veithen \latin{et~al.}(2005)Veithen, Gonze, and Ghosez]{Veithen_2005}
Veithen,~M.; Gonze,~X.; Ghosez,~P. Nonlinear optical susceptibilities, Raman
  efficiencies, and electro-optic tensors from first-principles density
  functional perturbation theory. \emph{Phys. Rev. B} \textbf{2005}, \emph{71},
  125107\relax
\mciteBstWouldAddEndPuncttrue
\mciteSetBstMidEndSepPunct{\mcitedefaultmidpunct}
{\mcitedefaultendpunct}{\mcitedefaultseppunct}\relax
\EndOfBibitem
\bibitem[Prussel and V\'eniard(2018)Prussel, and V\'eniard]{Prussel_2018}
Prussel,~L.; V\'eniard,~V. Linear electro-optic effect in semiconductors: Ab
  initio description of the electronic contribution. \emph{Phys. Rev. B}
  \textbf{2018}, \emph{97}, 205201\relax
\mciteBstWouldAddEndPuncttrue
\mciteSetBstMidEndSepPunct{\mcitedefaultmidpunct}
{\mcitedefaultendpunct}{\mcitedefaultseppunct}\relax
\EndOfBibitem
\bibitem[Friese \latin{et~al.}(2015)Friese, Beerepoot, Ringholm, and
  Ruud]{Friese2015_1}
Friese,~D.~H.; Beerepoot,~M. T.~P.; Ringholm,~M.; Ruud,~K. Open-Ended Recursive
  Approach for the Calculation of Multiphoton Absorption Matrix Elements.
  \emph{Journal of Chemical Theory and Computation} \textbf{2015}, \emph{11},
  1129--1144\relax
\mciteBstWouldAddEndPuncttrue
\mciteSetBstMidEndSepPunct{\mcitedefaultmidpunct}
{\mcitedefaultendpunct}{\mcitedefaultseppunct}\relax
\EndOfBibitem
\bibitem[Friese \latin{et~al.}(2015)Friese, Bast, and Ruud]{Friese2015_2}
Friese,~D.~H.; Bast,~R.; Ruud,~K. Five-Photon Absorption and Selective
  Enhancement of Multiphoton Absorption Processes. \emph{ACS Photonics}
  \textbf{2015}, \emph{2}, 572--577\relax
\mciteBstWouldAddEndPuncttrue
\mciteSetBstMidEndSepPunct{\mcitedefaultmidpunct}
{\mcitedefaultendpunct}{\mcitedefaultseppunct}\relax
\EndOfBibitem
\bibitem[Brabec and Krausz(2000)Brabec, and Krausz]{Brabec2000}
Brabec,~T.; Krausz,~F. Intense few-cycle laser fields: Frontiers of nonlinear
  optics. \emph{Rev. Mod. Phys.} \textbf{2000}, \emph{72}, 545--591\relax
\mciteBstWouldAddEndPuncttrue
\mciteSetBstMidEndSepPunct{\mcitedefaultmidpunct}
{\mcitedefaultendpunct}{\mcitedefaultseppunct}\relax
\EndOfBibitem
\bibitem[Lorin \latin{et~al.}(2015)Lorin, Lytova, Memarian, and
  Bandrauk]{Lorin2015}
Lorin,~E.; Lytova,~M.; Memarian,~A.; Bandrauk,~A.~D. Development of
  nonperturbative nonlinear optics models including effects of high order
  nonlinearities and of free electron plasma:
  Maxwell{\textendash}Schr{\"{o}}dinger equations coupled with evolution
  equations for polarization effects, and the {SFA}-like nonlinear optics
  model. \emph{Journal of Physics A: Mathematical and Theoretical}
  \textbf{2015}, \emph{48}, 105201\relax
\mciteBstWouldAddEndPuncttrue
\mciteSetBstMidEndSepPunct{\mcitedefaultmidpunct}
{\mcitedefaultendpunct}{\mcitedefaultseppunct}\relax
\EndOfBibitem
\bibitem[Peterson(1967)]{Peterson1967}
Peterson,~R.~L. {Formal Theory of Nonlinear Response}. \emph{Rev. Mod. Phys.}
  \textbf{1967}, \emph{39}, 69--77\relax
\mciteBstWouldAddEndPuncttrue
\mciteSetBstMidEndSepPunct{\mcitedefaultmidpunct}
{\mcitedefaultendpunct}{\mcitedefaultseppunct}\relax
\EndOfBibitem
\bibitem[Safi and Joyez(2011)Safi, and Joyez]{Safi2011}
Safi,~I.; Joyez,~P. Time-dependent theory of nonlinear response and current
  fluctuations. \emph{Phys. Rev. B} \textbf{2011}, \emph{84}, 205129\relax
\mciteBstWouldAddEndPuncttrue
\mciteSetBstMidEndSepPunct{\mcitedefaultmidpunct}
{\mcitedefaultendpunct}{\mcitedefaultseppunct}\relax
\EndOfBibitem
\bibitem[Strelkov(2016)]{Strelkov2016}
Strelkov,~V.~V. High-order optical processes in intense laser field: Towards
  nonperturbative nonlinear optics. \emph{Phys. Rev. A} \textbf{2016},
  \emph{93}, 053812\relax
\mciteBstWouldAddEndPuncttrue
\mciteSetBstMidEndSepPunct{\mcitedefaultmidpunct}
{\mcitedefaultendpunct}{\mcitedefaultseppunct}\relax
\EndOfBibitem
\bibitem[Goings \latin{et~al.}(2018)Goings, Lestrange, and Li]{Goings2018}
Goings,~J.~J.; Lestrange,~P.~J.; Li,~X. Real-time time-dependent electronic
  structure theory. \emph{Wiley Interdiscip. Rev. Comput. Mol. Sci.}
  \textbf{2018}, \emph{8}, 1--19\relax
\mciteBstWouldAddEndPuncttrue
\mciteSetBstMidEndSepPunct{\mcitedefaultmidpunct}
{\mcitedefaultendpunct}{\mcitedefaultseppunct}\relax
\EndOfBibitem
\bibitem[Provorse and Isborn(2016)Provorse, and Isborn]{Isoborn_2016}
Provorse,~M.~R.; Isborn,~C.~M. Electron dynamics with real-time time-dependent
  density functional theory. \emph{International Journal of Quantum Chemistry}
  \textbf{2016}, \emph{116}, 739--749\relax
\mciteBstWouldAddEndPuncttrue
\mciteSetBstMidEndSepPunct{\mcitedefaultmidpunct}
{\mcitedefaultendpunct}{\mcitedefaultseppunct}\relax
\EndOfBibitem
\bibitem[Cho \latin{et~al.}(2018)Cho, Rouxel, Kowalewski, Saurabh, Lee, and
  Mukamel]{Cho_2018}
Cho,~D.; Rouxel,~J.~R.; Kowalewski,~M.; Saurabh,~P.; Lee,~J.~Y.; Mukamel,~S.
  Phase Cycling RT-TDDFT Simulation Protocol for Nonlinear XUV and X-ray
  Molecular Spectroscopy. \emph{The Journal of Physical Chemistry Letters}
  \textbf{2018}, \emph{9}, 1072--1078\relax
\mciteBstWouldAddEndPuncttrue
\mciteSetBstMidEndSepPunct{\mcitedefaultmidpunct}
{\mcitedefaultendpunct}{\mcitedefaultseppunct}\relax
\EndOfBibitem
\bibitem[Ding \latin{et~al.}(2013)Ding, Van~Kuiken, Eichinger, and
  Li]{Ding_2013}
Ding,~F.; Van~Kuiken,~B.~E.; Eichinger,~B.~E.; Li,~X. An efficient method for
  calculating dynamical hyperpolarizabilities using real-time time-dependent
  density functional theory. \emph{The Journal of Chemical Physics}
  \textbf{2013}, \emph{138}, 064104\relax
\mciteBstWouldAddEndPuncttrue
\mciteSetBstMidEndSepPunct{\mcitedefaultmidpunct}
{\mcitedefaultendpunct}{\mcitedefaultseppunct}\relax
\EndOfBibitem
\bibitem[Mattiat and Luber(2018)Mattiat, and Luber]{Mattiat_2018}
Mattiat,~J.; Luber,~S. Efficient calculation of (resonance) Raman spectra and
  excitation profiles with real-time propagation. \emph{The Journal of Chemical
  Physics} \textbf{2018}, \emph{149}, 174108\relax
\mciteBstWouldAddEndPuncttrue
\mciteSetBstMidEndSepPunct{\mcitedefaultmidpunct}
{\mcitedefaultendpunct}{\mcitedefaultseppunct}\relax
\EndOfBibitem
\bibitem[Penka~Fowe and Bandrauk(2011)Penka~Fowe, and Bandrauk]{Bandrauk_2011}
Penka~Fowe,~E.; Bandrauk,~A.~D. Nonperturbative time-dependent
  density-functional theory of ionization and harmonic generation in OCS and
  CS${}_{2}$ molecules with ultrashort intense laser pulses: Intensity and
  orientational effects. \emph{Phys. Rev. A} \textbf{2011}, \emph{84},
  035402\relax
\mciteBstWouldAddEndPuncttrue
\mciteSetBstMidEndSepPunct{\mcitedefaultmidpunct}
{\mcitedefaultendpunct}{\mcitedefaultseppunct}\relax
\EndOfBibitem
\bibitem[Luppi and Head-Gordon(2012)Luppi, and Head-Gordon]{Luppi_2012}
Luppi,~E.; Head-Gordon,~M. Computation of high-harmonic generation spectra of
  H2 and N2 in intense laser pulses using quantum chemistry methods and
  time-dependent density functional theory. \emph{Molecular Physics}
  \textbf{2012}, \emph{110}, 909--923\relax
\mciteBstWouldAddEndPuncttrue
\mciteSetBstMidEndSepPunct{\mcitedefaultmidpunct}
{\mcitedefaultendpunct}{\mcitedefaultseppunct}\relax
\EndOfBibitem
\bibitem[Nguyen \latin{et~al.}(2016)Nguyen, Koh, Lefelhocz, and
  Parkhill]{Nguyen_2016}
Nguyen,~T.~S.; Koh,~J.~H.; Lefelhocz,~S.; Parkhill,~J. Black-Box, Real-Time
  Simulations of Transient Absorption Spectroscopy. \emph{The Journal of
  Physical Chemistry Letters} \textbf{2016}, \emph{7}, 1590--1595\relax
\mciteBstWouldAddEndPuncttrue
\mciteSetBstMidEndSepPunct{\mcitedefaultmidpunct}
{\mcitedefaultendpunct}{\mcitedefaultseppunct}\relax
\EndOfBibitem
\bibitem[Tancogne-Dejean \latin{et~al.}(2017)Tancogne-Dejean, M\"ucke,
  K\"artner, and Rubio]{Tancogne-Dejean_2017}
Tancogne-Dejean,~N.; M\"ucke,~O.~D.; K\"artner,~F.~X.; Rubio,~A. Impact of the
  Electronic Band Structure in High-Harmonic Generation Spectra of Solids.
  \emph{Phys. Rev. Lett.} \textbf{2017}, \emph{118}, 087403\relax
\mciteBstWouldAddEndPuncttrue
\mciteSetBstMidEndSepPunct{\mcitedefaultmidpunct}
{\mcitedefaultendpunct}{\mcitedefaultseppunct}\relax
\EndOfBibitem
\bibitem[Sato \latin{et~al.}(2015)Sato, Yabana, Shinohara, Otobe, Lee, and
  Bertsch]{Yabana_2015}
Sato,~S.~A.; Yabana,~K.; Shinohara,~Y.; Otobe,~T.; Lee,~K.-M.; Bertsch,~G.~F.
  Time-dependent density functional theory of high-intensity short-pulse laser
  irradiation on insulators. \emph{Phys. Rev. B} \textbf{2015}, \emph{92},
  205413\relax
\mciteBstWouldAddEndPuncttrue
\mciteSetBstMidEndSepPunct{\mcitedefaultmidpunct}
{\mcitedefaultendpunct}{\mcitedefaultseppunct}\relax
\EndOfBibitem
\bibitem[Ullrich(2011)]{Ullrich2011}
Ullrich,~C.~A. \emph{Time-dependent density-functional theory: concepts and
  applications}; Oxford University Press: Oxford, 2011\relax
\mciteBstWouldAddEndPuncttrue
\mciteSetBstMidEndSepPunct{\mcitedefaultmidpunct}
{\mcitedefaultendpunct}{\mcitedefaultseppunct}\relax
\EndOfBibitem
\bibitem[Marques \latin{et~al.}(2012)Marques, Maitra, Nogueira, Gross, and
  Rubio]{Marques2012}
Marques,~M.~A.; Maitra,~N.~T.; Nogueira,~F.~M.; Gross,~E. K.~U.; Rubio,~A.
  \emph{Fundamentals of Time-Dependent Density Functional Theory}; Lecture
  Notes in Physics book series; Springer, Berlin, Heidelberg, 2012\relax
\mciteBstWouldAddEndPuncttrue
\mciteSetBstMidEndSepPunct{\mcitedefaultmidpunct}
{\mcitedefaultendpunct}{\mcitedefaultseppunct}\relax
\EndOfBibitem
\bibitem[Takimoto \latin{et~al.}(2007)Takimoto, Vila, and Rehr]{Takimoto2007}
Takimoto,~Y.; Vila,~F.~D.; Rehr,~J.~J. Real-time time-dependent density
  functional theory approach for frequency-dependent nonlinear optical response
  in photonic molecules. \emph{J. Chem. Phys.} \textbf{2007}, \emph{127}\relax
\mciteBstWouldAddEndPuncttrue
\mciteSetBstMidEndSepPunct{\mcitedefaultmidpunct}
{\mcitedefaultendpunct}{\mcitedefaultseppunct}\relax
\EndOfBibitem
\bibitem[Attaccalite and Gr{\"{u}}ning(2013)Attaccalite, and
  Gr{\"{u}}ning]{Attaccalite2013}
Attaccalite,~C.; Gr{\"{u}}ning,~M. Nonlinear optics from an \textit{ab initio}
  approach by means of the dynamical Berry phase: Application to second- and
  third-harmonic generation in semiconductors. \emph{Phys. Rev. B}
  \textbf{2013}, \emph{88}, 1--10\relax
\mciteBstWouldAddEndPuncttrue
\mciteSetBstMidEndSepPunct{\mcitedefaultmidpunct}
{\mcitedefaultendpunct}{\mcitedefaultseppunct}\relax
\EndOfBibitem
\bibitem[Uemoto \latin{et~al.}(2019)Uemoto, Kuwabara, Sato, and
  Yabana]{Uemoto2018}
Uemoto,~M.; Kuwabara,~Y.; Sato,~S.~A.; Yabana,~K. Nonlinear polarization
  evolution using time-dependent density functional theory. \emph{J. Chem.
  Phys.} \textbf{2019}, \emph{150}, 094101\relax
\mciteBstWouldAddEndPuncttrue
\mciteSetBstMidEndSepPunct{\mcitedefaultmidpunct}
{\mcitedefaultendpunct}{\mcitedefaultseppunct}\relax
\EndOfBibitem
\bibitem[Luppi and Head-Gordon(2012)Luppi, and Head-Gordon]{Luppi2012}
Luppi,~E.; Head-Gordon,~M. Computation of high-harmonic generation spectra of
  H2 and N2 in intense laser pulses using quantum chemistry methods and
  time-dependent density functional theory. \emph{Molecular Physics}
  \textbf{2012}, \emph{110}, 909--923\relax
\mciteBstWouldAddEndPuncttrue
\mciteSetBstMidEndSepPunct{\mcitedefaultmidpunct}
{\mcitedefaultendpunct}{\mcitedefaultseppunct}\relax
\EndOfBibitem
\bibitem[Cocchi \latin{et~al.}(2014)Cocchi, Prezzi, Ruini, Molinari, and
  Rozzi]{Cocchi2014}
Cocchi,~C.; Prezzi,~D.; Ruini,~A.; Molinari,~E.; Rozzi,~C.~A. \textit{Ab
  initio} simulation of optical limiting: The case of metal-free
  phthalocyanine. \emph{Phys. Rev. Lett.} \textbf{2014}, \emph{112}, 1--5\relax
\mciteBstWouldAddEndPuncttrue
\mciteSetBstMidEndSepPunct{\mcitedefaultmidpunct}
{\mcitedefaultendpunct}{\mcitedefaultseppunct}\relax
\EndOfBibitem
\bibitem[Alonso \latin{et~al.}(2008)Alonso, Andrade, Echenique, Falceto,
  Prada-Gracia, and Rubio]{Alonso2008}
Alonso,~J.~L.; Andrade,~X.; Echenique,~P.; Falceto,~F.; Prada-Gracia,~D.;
  Rubio,~A. {Efficient Formalism for Large-Scale Ab Initio Molecular Dynamics
  based on Time-Dependent Density Functional Theory}. \emph{Physical Review
  Letters} \textbf{2008}, \emph{101}, 096403\relax
\mciteBstWouldAddEndPuncttrue
\mciteSetBstMidEndSepPunct{\mcitedefaultmidpunct}
{\mcitedefaultendpunct}{\mcitedefaultseppunct}\relax
\EndOfBibitem
\bibitem[Falke \latin{et~al.}(2014)Falke, Rozzi, Brida, Maiuri, Amato, Sommer,
  De~Sio, Rubio, Cerullo, Molinari, and Lienau]{Falke2014}
Falke,~S.~M.; Rozzi,~C.~A.; Brida,~D.; Maiuri,~M.; Amato,~M.; Sommer,~E.;
  De~Sio,~A.; Rubio,~A.; Cerullo,~G.; Molinari,~E.; Lienau,~C. Coherent
  ultrafast charge transfer in an organic photovoltaic blend. \emph{Science}
  \textbf{2014}, \emph{344}, 1001--1005\relax
\mciteBstWouldAddEndPuncttrue
\mciteSetBstMidEndSepPunct{\mcitedefaultmidpunct}
{\mcitedefaultendpunct}{\mcitedefaultseppunct}\relax
\EndOfBibitem
\bibitem[Pittalis \latin{et~al.}(2015)Pittalis, Delgado, Robin, Freimuth,
  Christoffers, Lienau, and Rozzi]{Pittalis2015}
Pittalis,~S.; Delgado,~A.; Robin,~J.; Freimuth,~L.; Christoffers,~J.;
  Lienau,~C.; Rozzi,~C.~A. {Charge Separation Dynamics and Opto-Electronic
  Properties of a Diaminoterephthalate-C 60 Dyad}. \emph{Advanced Functional
  Materials} \textbf{2015}, \emph{25}, 2047--2053\relax
\mciteBstWouldAddEndPuncttrue
\mciteSetBstMidEndSepPunct{\mcitedefaultmidpunct}
{\mcitedefaultendpunct}{\mcitedefaultseppunct}\relax
\EndOfBibitem
\bibitem[Rozzi \latin{et~al.}(2017)Rozzi, Troiani, and Tavernelli]{Rozzi_2017}
Rozzi,~C.~A.; Troiani,~F.; Tavernelli,~I. Quantum modeling of ultrafast
  photoinduced charge separation. \emph{J.~Phys.:~Condens.~Matter}
  \textbf{2017}, \emph{30}, 013002\relax
\mciteBstWouldAddEndPuncttrue
\mciteSetBstMidEndSepPunct{\mcitedefaultmidpunct}
{\mcitedefaultendpunct}{\mcitedefaultseppunct}\relax
\EndOfBibitem
\bibitem[Rozzi and Pittalis(2018)Rozzi, and Pittalis]{Rozzi2018}
Rozzi,~C.~A.; Pittalis,~S. \emph{Handbook of Materials Modeling}; Springer
  International Publishing: Cham, 2018; pp 1--19\relax
\mciteBstWouldAddEndPuncttrue
\mciteSetBstMidEndSepPunct{\mcitedefaultmidpunct}
{\mcitedefaultendpunct}{\mcitedefaultseppunct}\relax
\EndOfBibitem
\bibitem[Yamada and Yabana(2019)Yamada, and Yabana]{Yamada2019}
Yamada,~A.; Yabana,~K. Multiscale time-dependent density functional theory for
  a unified description of ultrafast dynamics: Pulsed light, electron, and
  lattice motions in crystalline solids. \emph{Phys. Rev. B} \textbf{2019},
  \emph{99}, 245103\relax
\mciteBstWouldAddEndPuncttrue
\mciteSetBstMidEndSepPunct{\mcitedefaultmidpunct}
{\mcitedefaultendpunct}{\mcitedefaultseppunct}\relax
\EndOfBibitem
\bibitem[Jacobs \latin{et~al.}(2020)Jacobs, Krumland, Valencia, Wang, Rossi,
  and Cocchi]{Jacobs2020}
Jacobs,~M.; Krumland,~J.; Valencia,~A.~M.; Wang,~H.; Rossi,~M.; Cocchi,~C.
  Ultrafast charge transfer and vibronic coupling in a laser-excited hybrid
  inorganic/organic interface. \emph{Advances in Physics: X} \textbf{2020},
  \emph{5}, 1749883\relax
\mciteBstWouldAddEndPuncttrue
\mciteSetBstMidEndSepPunct{\mcitedefaultmidpunct}
{\mcitedefaultendpunct}{\mcitedefaultseppunct}\relax
\EndOfBibitem
\bibitem[Krumland \latin{et~al.}(2020)Krumland, Valencia, Pittalis, Rozzi, and
  Cocchi]{krumland2020}
Krumland,~J.; Valencia,~A.~M.; Pittalis,~S.; Rozzi,~C.~A.; Cocchi,~C.
  Understanding real-time time-dependent density-functional theory simulations
  of ultrafast laser-induced dynamics in organic molecules. \emph{arXiv
  preprint arXiv:2003.08669} \textbf{2020}, \relax
\mciteBstWouldAddEndPunctfalse
\mciteSetBstMidEndSepPunct{\mcitedefaultmidpunct}
{}{\mcitedefaultseppunct}\relax
\EndOfBibitem
\bibitem[Yuen-Zhou \latin{et~al.}(2010)Yuen-Zhou, Tempel,
  Rodr\'{\i}guez-Rosario, and Aspuru-Guzik]{YuenZhou2010}
Yuen-Zhou,~J.; Tempel,~D.~G.; Rodr\'{\i}guez-Rosario,~C.~A.; Aspuru-Guzik,~A.
  Time-Dependent Density Functional Theory for Open Quantum Systems with
  Unitary Propagation. \emph{Phys. Rev. Lett.} \textbf{2010}, \emph{104},
  043001\relax
\mciteBstWouldAddEndPuncttrue
\mciteSetBstMidEndSepPunct{\mcitedefaultmidpunct}
{\mcitedefaultendpunct}{\mcitedefaultseppunct}\relax
\EndOfBibitem
\bibitem[{Newcomb}(1963)]{Newcomb1963}
{Newcomb},~R.~W. Distributional impulse response theorems. \emph{Proceedings of
  the IEEE} \textbf{1963}, \emph{51}, 1157--1158\relax
\mciteBstWouldAddEndPuncttrue
\mciteSetBstMidEndSepPunct{\mcitedefaultmidpunct}
{\mcitedefaultendpunct}{\mcitedefaultseppunct}\relax
\EndOfBibitem
\bibitem[Yabana and Bertsch(1996)Yabana, and Bertsch]{Yabana1996}
Yabana,~K.; Bertsch,~G.~F. Time-dependent local-density approximation in real
  time. \emph{Phys. Rev. B} \textbf{1996}, \emph{54}, 4484--4487\relax
\mciteBstWouldAddEndPuncttrue
\mciteSetBstMidEndSepPunct{\mcitedefaultmidpunct}
{\mcitedefaultendpunct}{\mcitedefaultseppunct}\relax
\EndOfBibitem
\bibitem[Dini \latin{et~al.}(2016)Dini, Calvete, and Hanack]{Dini2016}
Dini,~D.; Calvete,~M.~J.; Hanack,~M. Nonlinear Optical Materials for the Smart
  Filtering of Optical Radiation. \emph{Chem. Rev.} \textbf{2016}, \emph{116},
  13043--13233\relax
\mciteBstWouldAddEndPuncttrue
\mciteSetBstMidEndSepPunct{\mcitedefaultmidpunct}
{\mcitedefaultendpunct}{\mcitedefaultseppunct}\relax
\EndOfBibitem
\bibitem[Sun and Riggs(1999)Sun, and Riggs]{Yaping2019}
Sun,~Y.-P.; Riggs,~J.~E. Organic and inorganic optical limiting materials.
  {F}rom fullerenes to nanoparticles. \emph{Int. Rev. Phys. Chem.}
  \textbf{1999}, \emph{18}, 43--90\relax
\mciteBstWouldAddEndPuncttrue
\mciteSetBstMidEndSepPunct{\mcitedefaultmidpunct}
{\mcitedefaultendpunct}{\mcitedefaultseppunct}\relax
\EndOfBibitem
\bibitem[Miao \latin{et~al.}(2019)Miao, Sang, Song, Liang, Liu, Sun, and
  Xu]{Miao2019}
Miao,~Q.; Sang,~Z.; Song,~R.; Liang,~M.; Liu,~Q.; Sun,~E.; Xu,~Y. Nonlinear
  properties of chloroindium phthalocyanines with nanosecond pulses.
  \emph{Journal of Photochemistry and Photobiology A: Chemistry} \textbf{2019},
  \emph{385}, 112087\relax
\mciteBstWouldAddEndPuncttrue
\mciteSetBstMidEndSepPunct{\mcitedefaultmidpunct}
{\mcitedefaultendpunct}{\mcitedefaultseppunct}\relax
\EndOfBibitem
\bibitem[de~Wergifosse and Grimme(2019)de~Wergifosse, and Grimme]{Grimme_2019}
de~Wergifosse,~M.; Grimme,~S. Nonlinear-response properties in a simplified
  time-dependent density functional theory (sTD-DFT) framework: Evaluation of
  excited-state absorption spectra. \emph{The Journal of Chemical Physics}
  \textbf{2019}, \emph{150}, 094112\relax
\mciteBstWouldAddEndPuncttrue
\mciteSetBstMidEndSepPunct{\mcitedefaultmidpunct}
{\mcitedefaultendpunct}{\mcitedefaultseppunct}\relax
\EndOfBibitem
\bibitem[Fischer \latin{et~al.}(2015)Fischer, Cramer, and Govind]{Govind_2015}
Fischer,~S.~A.; Cramer,~C.~J.; Govind,~N. Excited State Absorption from
  Real-Time Time-Dependent Density Functional Theory. \emph{Journal of Chemical
  Theory and Computation} \textbf{2015}, \emph{11}, 4294--4303\relax
\mciteBstWouldAddEndPuncttrue
\mciteSetBstMidEndSepPunct{\mcitedefaultmidpunct}
{\mcitedefaultendpunct}{\mcitedefaultseppunct}\relax
\EndOfBibitem
\bibitem[Fischer \latin{et~al.}(2016)Fischer, Cramer, and Govind]{Govind_2016}
Fischer,~S.~A.; Cramer,~C.~J.; Govind,~N. Excited-State Absorption from
  Real-Time Time-Dependent Density Functional Theory: Optical Limiting in Zinc
  Phthalocyanine. \emph{The Journal of Physical Chemistry Letters}
  \textbf{2016}, \emph{7}, 1387--1391\relax
\mciteBstWouldAddEndPuncttrue
\mciteSetBstMidEndSepPunct{\mcitedefaultmidpunct}
{\mcitedefaultendpunct}{\mcitedefaultseppunct}\relax
\EndOfBibitem
\bibitem[Bowman \latin{et~al.}(2017)Bowman, Asher, Fischer, Cramer, and
  Govind]{Govind_2017}
Bowman,~D.~N.; Asher,~J.~C.; Fischer,~S.~A.; Cramer,~C.~J.; Govind,~N.
  Excited-state absorption in tetrapyridyl porphyrins: comparing real-time and
  quadratic-response time-dependent density functional theory. \emph{Phys.
  Chem. Chem. Phys.} \textbf{2017}, \emph{19}, 27452--27462\relax
\mciteBstWouldAddEndPuncttrue
\mciteSetBstMidEndSepPunct{\mcitedefaultmidpunct}
{\mcitedefaultendpunct}{\mcitedefaultseppunct}\relax
\EndOfBibitem
\bibitem[Ghosh \latin{et~al.}(2019)Ghosh, Asher, Gagliardi, Cramer, and
  Govind]{Govind_2019}
Ghosh,~S.; Asher,~J.~C.; Gagliardi,~L.; Cramer,~C.~J.; Govind,~N. A
  semiempirical effective Hamiltonian based approach for analyzing excited
  state wave functions and computing excited state absorption spectra using
  real-time dynamics. \emph{The Journal of Chemical Physics} \textbf{2019},
  \emph{150}, 104103\relax
\mciteBstWouldAddEndPuncttrue
\mciteSetBstMidEndSepPunct{\mcitedefaultmidpunct}
{\mcitedefaultendpunct}{\mcitedefaultseppunct}\relax
\EndOfBibitem
\bibitem[Elliott and Maitra(2012)Elliott, and Maitra]{Maitra_2012}
Elliott,~P.; Maitra,~N.~T. Propagation of initially excited states in
  time-dependent density-functional theory. \emph{Phys. Rev. A} \textbf{2012},
  \emph{85}, 052510\relax
\mciteBstWouldAddEndPuncttrue
\mciteSetBstMidEndSepPunct{\mcitedefaultmidpunct}
{\mcitedefaultendpunct}{\mcitedefaultseppunct}\relax
\EndOfBibitem
\bibitem[Mosquera \latin{et~al.}(2016)Mosquera, Chen, Ratner, and
  Schatz]{Mosquera_2016}
Mosquera,~M.~A.; Chen,~L.~X.; Ratner,~M.~A.; Schatz,~G.~C. Sequential double
  excitations from linear-response time-dependent density functional theory.
  \emph{The Journal of Chemical Physics} \textbf{2016}, \emph{144},
  204105\relax
\mciteBstWouldAddEndPuncttrue
\mciteSetBstMidEndSepPunct{\mcitedefaultmidpunct}
{\mcitedefaultendpunct}{\mcitedefaultseppunct}\relax
\EndOfBibitem
\bibitem[Sheng \latin{et~al.}(2020)Sheng, Zhu, Yin, Chen, Wang, Wang, Shao, and
  Chen]{Sheng_2020}
Sheng,~X.; Zhu,~H.; Yin,~K.; Chen,~J.; Wang,~J.; Wang,~C.; Shao,~J.; Chen,~F.
  Excited-State Absorption by Linear Response Time-Dependent Density Functional
  Theory. \emph{The Journal of Physical Chemistry C} \textbf{2020}, \emph{124},
  4693--4700\relax
\mciteBstWouldAddEndPuncttrue
\mciteSetBstMidEndSepPunct{\mcitedefaultmidpunct}
{\mcitedefaultendpunct}{\mcitedefaultseppunct}\relax
\EndOfBibitem
\bibitem[Bellier \latin{et~al.}(2012)Bellier, Makarov, Bouit, Rigaut, Kamada,
  Feneyrou, Berginc, Maury, Perry, and Andraud]{Bellier2012}
Bellier,~Q.; Makarov,~N.~S.; Bouit,~P.-A.; Rigaut,~S.; Kamada,~K.;
  Feneyrou,~P.; Berginc,~G.; Maury,~O.; Perry,~J.~W.; Andraud,~C. Excited state
  absorption: a key phenomenon for the improvement of biphotonic based optical
  limiting at telecommunication wavelengths. \emph{Phys. Chem. Chem. Phys.}
  \textbf{2012}, \emph{14}, 15299--15307\relax
\mciteBstWouldAddEndPuncttrue
\mciteSetBstMidEndSepPunct{\mcitedefaultmidpunct}
{\mcitedefaultendpunct}{\mcitedefaultseppunct}\relax
\EndOfBibitem
\bibitem[Fischer \latin{et~al.}(2016)Fischer, Cramer, and Govind]{Fischer2016}
Fischer,~S.~A.; Cramer,~C.~J.; Govind,~N. {Excited-State Absorption from
  Real-Time Time-Dependent Density Functional Theory: Optical Limiting in Zinc
  Phthalocyanine}. \emph{The Journal of Physical Chemistry Letters}
  \textbf{2016}, \emph{7}, 1387--1391\relax
\mciteBstWouldAddEndPuncttrue
\mciteSetBstMidEndSepPunct{\mcitedefaultmidpunct}
{\mcitedefaultendpunct}{\mcitedefaultseppunct}\relax
\EndOfBibitem
\bibitem[Fuks and Maitra(2014)Fuks, and Maitra]{Neepa2014}
Fuks,~J.~I.; Maitra,~N.~T. Challenging adiabatic time-dependent density
  functional theory with a Hubbard dimer: the case of time-resolved long-range
  charge transfer. \emph{Phys. Chem. Chem. Phys.} \textbf{2014}, \emph{16},
  14504--14513\relax
\mciteBstWouldAddEndPuncttrue
\mciteSetBstMidEndSepPunct{\mcitedefaultmidpunct}
{\mcitedefaultendpunct}{\mcitedefaultseppunct}\relax
\EndOfBibitem
\bibitem[Luo \latin{et~al.}(2013)Luo, Elliott, and Maitra]{Neepa2013a}
Luo,~K.; Elliott,~P.; Maitra,~N.~T. Absence of dynamical steps in the exact
  correlation potential in the linear response regime. \emph{Phys. Rev. A}
  \textbf{2013}, \emph{88}, 042508\relax
\mciteBstWouldAddEndPuncttrue
\mciteSetBstMidEndSepPunct{\mcitedefaultmidpunct}
{\mcitedefaultendpunct}{\mcitedefaultseppunct}\relax
\EndOfBibitem
\bibitem[Fuks \latin{et~al.}(2013)Fuks, Elliott, Rubio, and Maitra]{Neepa2013b}
Fuks,~J.~I.; Elliott,~P.; Rubio,~A.; Maitra,~N.~T. Dynamics of Charge-Transfer
  Processes with Time-Dependent Density Functional Theory. \emph{The Journal of
  Physical Chemistry Letters} \textbf{2013}, \emph{4}, 735--739, PMID:
  26281927\relax
\mciteBstWouldAddEndPuncttrue
\mciteSetBstMidEndSepPunct{\mcitedefaultmidpunct}
{\mcitedefaultendpunct}{\mcitedefaultseppunct}\relax
\EndOfBibitem
\bibitem[Elliott \latin{et~al.}(2012)Elliott, Fuks, Rubio, and
  Maitra]{Neepa2012}
Elliott,~P.; Fuks,~J.~I.; Rubio,~A.; Maitra,~N.~T. Universal Dynamical Steps in
  the Exact Time-Dependent Exchange-Correlation Potential. \emph{Phys. Rev.
  Lett.} \textbf{2012}, \emph{109}, 266404\relax
\mciteBstWouldAddEndPuncttrue
\mciteSetBstMidEndSepPunct{\mcitedefaultmidpunct}
{\mcitedefaultendpunct}{\mcitedefaultseppunct}\relax
\EndOfBibitem
\bibitem[Maitra(2005)]{Neepa2005}
Maitra,~N.~T. Undoing static correlation: Long-range charge transfer in
  time-dependent density-functional theory. \emph{The Journal of Chemical
  Physics} \textbf{2005}, \emph{122}, 234104\relax
\mciteBstWouldAddEndPuncttrue
\mciteSetBstMidEndSepPunct{\mcitedefaultmidpunct}
{\mcitedefaultendpunct}{\mcitedefaultseppunct}\relax
\EndOfBibitem
\bibitem[Mardirossian and Head-Gordon(2017)Mardirossian, and
  Head-Gordon]{Head-Gordon2017}
Mardirossian,~N.; Head-Gordon,~M. Thirty years of density functional theory in
  computational chemistry: an overview and extensive assessment of 200 density
  functionals. \emph{Molecular Physics} \textbf{2017}, \emph{115},
  2315--2372\relax
\mciteBstWouldAddEndPuncttrue
\mciteSetBstMidEndSepPunct{\mcitedefaultmidpunct}
{\mcitedefaultendpunct}{\mcitedefaultseppunct}\relax
\EndOfBibitem
\bibitem[Perfetto and Stefanucci(2015)Perfetto, and Stefanucci]{Perfetto2015}
Perfetto,~E.; Stefanucci,~G. Some exact properties of the nonequilibrium
  response function for transient photoabsorption. \emph{Phys. Rev. A}
  \textbf{2015}, \emph{91}, 033416\relax
\mciteBstWouldAddEndPuncttrue
\mciteSetBstMidEndSepPunct{\mcitedefaultmidpunct}
{\mcitedefaultendpunct}{\mcitedefaultseppunct}\relax
\EndOfBibitem
\bibitem[Santhi \latin{et~al.}(2006)Santhi, Namboodiri, Radhakrishnan, and
  Nampoori]{Santhi2006}
Santhi,~A.; Namboodiri,~V.~V.; Radhakrishnan,~P.; Nampoori,~V. P.~N. {Spectral
  dependence of third order nonlinear optical susceptibility of zinc
  phthalocyanine}. \emph{J. Appl. Phys.} \textbf{2006}, \emph{100},
  053109\relax
\mciteBstWouldAddEndPuncttrue
\mciteSetBstMidEndSepPunct{\mcitedefaultmidpunct}
{\mcitedefaultendpunct}{\mcitedefaultseppunct}\relax
\EndOfBibitem
\bibitem[Marian(2012)]{marian2012spin}
Marian,~C.~M. Spin--orbit coupling and intersystem crossing in molecules.
  \emph{Wiley Interdiscip. Rev. Comput. Mol. Sci.} \textbf{2012}, \emph{2},
  187--203\relax
\mciteBstWouldAddEndPuncttrue
\mciteSetBstMidEndSepPunct{\mcitedefaultmidpunct}
{\mcitedefaultendpunct}{\mcitedefaultseppunct}\relax
\EndOfBibitem
\bibitem[Oliveira \latin{et~al.}(2008)Oliveira, Castro, Marques, and
  Rubio]{Oliveira2008}
Oliveira,~M. J.~T.; Castro,~A.; Marques,~M. A.~L.; Rubio,~A. {On the Use of
  Neumann's Principle for the Calculation of the Polarizability Tensor of
  Nanostructures}. \emph{Journal of Nanoscience and Nanotechnology}
  \textbf{2008}, \emph{8}, 3392--3398\relax
\mciteBstWouldAddEndPuncttrue
\mciteSetBstMidEndSepPunct{\mcitedefaultmidpunct}
{\mcitedefaultendpunct}{\mcitedefaultseppunct}\relax
\EndOfBibitem
\bibitem[Tancogne-Dejean \latin{et~al.}(2020)Tancogne-Dejean, Eich, and
  Rubio]{magnons}
Tancogne-Dejean,~N.; Eich,~F.~G.; Rubio,~A. Time-Dependent Magnons from First
  Principles. \emph{Journal of Chemical Theory and Computation} \textbf{2020},
  \emph{16}, 1007--1017, PMID: 31922758\relax
\mciteBstWouldAddEndPuncttrue
\mciteSetBstMidEndSepPunct{\mcitedefaultmidpunct}
{\mcitedefaultendpunct}{\mcitedefaultseppunct}\relax
\EndOfBibitem
\bibitem[Su and Eberly(1991)Su, and Eberly]{Su1991}
Su,~Q.; Eberly,~J.~H. Model atom for multiphoton physics. \emph{Phys. Rev. A}
  \textbf{1991}, \emph{44}, 5997--6008\relax
\mciteBstWouldAddEndPuncttrue
\mciteSetBstMidEndSepPunct{\mcitedefaultmidpunct}
{\mcitedefaultendpunct}{\mcitedefaultseppunct}\relax
\EndOfBibitem
\bibitem[Tancogne-Dejean \latin{et~al.}(2020)Tancogne-Dejean, Oliveira,
  Andrade, Appel, Borca, Le~Breton, Buchholz, Castro, Corni, Correa,
  De~Giovannini, Delgado, Eich, Flick, Gil, Gomez, Helbig, H{\"u}bener,
  Jest{\"a}dt, Jornet-Somoza, Larsen, Lebedeva, L{\"u}ders, Marques, Ohlmann,
  Pipolo, Rampp, Rozzi, Strubbe, Sato, Sch{\"a}fer, Theophilou, Welden, and
  Rubio]{Octopus1}
Tancogne-Dejean,~N.; Oliveira,~M. J.~T.; Andrade,~X.; Appel,~H.; Borca,~C.~H.;
  Le~Breton,~G.; Buchholz,~F.; Castro,~A.; Corni,~S.; Correa,~A.~A.;
  De~Giovannini,~U.; Delgado,~A.; Eich,~F.~G.; Flick,~J.; Gil,~G.; Gomez,~A.;
  Helbig,~N.; H{\"u}bener,~H.; Jest{\"a}dt,~R.; Jornet-Somoza,~J.;
  Larsen,~A.~H.; Lebedeva,~I.~V.; L{\"u}ders,~M.; Marques,~M. A.~L.;
  Ohlmann,~S.~T.; Pipolo,~S.; Rampp,~M.; Rozzi,~C.~A.; Strubbe,~D.~A.;
  Sato,~S.~A.; Sch{\"a}fer,~C.; Theophilou,~I.; Welden,~A.; Rubio,~A. Octopus,
  a computational framework for exploring light-driven phenomena and quantum
  dynamics in extended and finite systems. \emph{The Journal of Chemical
  Physics} \textbf{2020}, \emph{152}, 124119\relax
\mciteBstWouldAddEndPuncttrue
\mciteSetBstMidEndSepPunct{\mcitedefaultmidpunct}
{\mcitedefaultendpunct}{\mcitedefaultseppunct}\relax
\EndOfBibitem
\bibitem[num()]{numdetails}
The simulations were performed on a regular spatial grid with a spacing of
  $0.015$ bohr and $L = 5.0$ bohr. The dipole impulsive perturbation is
  $\hat{H}'(t) = -\hat{d}\kappa\delta(t)$, where $\hat{d} =
  \hat{x}_1+\hat{x}_2$. After the impulse is applied, the wavefunction is
  propagated up to $150$ Ha$^{-1}$, employing a time step of $0.002$
  Ha$^{-1}$.\relax
\mciteBstWouldAddEndPunctfalse
\mciteSetBstMidEndSepPunct{\mcitedefaultmidpunct}
{}{\mcitedefaultseppunct}\relax
\EndOfBibitem
\bibitem[Jackson(1999)]{Jackson}
Jackson,~J.~D. \emph{Classical electrodynamics}, 3rd ed.; Wiley: New York,
  {NY}, 1999\relax
\mciteBstWouldAddEndPuncttrue
\mciteSetBstMidEndSepPunct{\mcitedefaultmidpunct}
{\mcitedefaultendpunct}{\mcitedefaultseppunct}\relax
\EndOfBibitem
\end{mcitethebibliography}

\end{document}


\maketitle

\numberwithin{equation}{section}
\setcounter{equation}{0}

\newpage

\section{Absorption cross section in the dipole approximation}\label{SecCrossSectAppx}

In the interaction between matter and an external transverse electric field, the absorption cross section at each frequency $\omega$ is defined as the ratio between the energy exchanged with the field during the interaction, $E_{exc}$, and the total field energy per unit area, $I_{in}$, 
\begin{equation}\label{EqCrSecDeff}
\sigma(\omega) = \frac{E_{exc}(\omega)}{I_{in}(\omega)} \ .
\end{equation}
In the dipole approximation, with no magnetic field applied, the light-matter interaction Hamiltonian is $\hat{H}(t) = \hat{d}_{\mu}\mathcal{E}_{\mu}(t)$, where $\hat{d}_{\mu}$ is the dipole operator. The total amount of energy exchanged during the interaction is
\begin{equation}\label{DefEabst}
\Delta E = \int\limits_{-\infty}^{+\infty}\frac{dE(t)}{dt}\ dt=\int\limits_{-\infty}^{+\infty} \frac{d }{dt} \mel{\Psi(t)}{\hat{H}(t)}{\Psi(t)}\ dt \ ,
\end{equation}
where $E(t)= \mel{\Psi(t)}{\hat{H}(t)}{\Psi(t)}$ is the total energy at time $t$.
By applying the Ehrenfest theorem to the integrand of Eq. \eqref{DefEabst}, we obtain
\begin{equation}\label{Erhth}
\frac{d }{dt} \mel{\Psi(t)}{\hat{H}(t)}{\Psi(t)} = \left\langle \Psi(t)\left|\frac{\partial \hat{H}(t)}{\partial t}\right|\Psi(t)\right\rangle = -d_{\mu}(t) \frac{d \mathcal{E}_{\mu}(t)}{d t} \ .
\end{equation}
Substituting Eq. \eqref{Erhth} into Eq. \eqref{DefEabst}, we have
\begin{equation}
\Delta E = -\int\limits_{-\infty}^{+\infty} dt \ d_{\mu}(t)  \frac{d\mathcal{E}_{\mu}(t)}{d t}.
\end{equation}
By using the Plancherel theorem, we change from time to the frequency domain
\begin{equation}\label{DefEabso2}
\Delta E = -\frac{i}{2\pi}\int\limits_{-\infty}^{+\infty} d\omega \ \omega\tilde{d}_{\mu}(\omega) \tilde{\mathcal{E}}^{*}_{\mu}(\omega),
\end{equation}
where $\tilde{d}_{\mu}(\omega)$ and $\tilde{\mathcal{E}}_{\mu}(\omega)$ are the Fourier transforms of $d_{\mu}(t)$ and $\mathcal{E}^{\mu}(t)$, respectively.
In writing Eq. \eqref{DefEabso2}, we have  also used the property of Fourier transform: $\mathcal{F}[df(t)/dt] = -i\omega\tilde{f}(\omega)$ (where $\mathcal{F}[\ldots]$ is the Fourier transform operator).

Since both $d_{\mu}(t)$ and $\mathcal{E}_{\mu}(t)$ are real quantities, their complex conjugates fulfill the relations $\tilde{d}_{\mu}(-\omega)=\tilde{d}^{*}_{\mu}(\omega)$ and $\tilde{\mathcal{E}}_{\mu}(-\omega)=\tilde{\mathcal{E}}^{*}_{\mu}(\omega)$.
Thus, we get
\begin{equation}\label{DefEabso}
\Delta E = \int\limits^{\infty}_{0} E_{exc}(\omega)  \ d\omega \ 
\end{equation}
where 
\begin{equation}\label{Eqdeomega}
E_{exc}(\omega) = \frac{1}{\pi}\omega\Im\left[\tilde{d}_{\mu}(\omega) \tilde{\mathcal{E}}^{*}_{\mu}(\omega)\right]
\end{equation}
may be regarded as the energy absorbed at each frequency $\omega$ --- a quantity which is defined for  $\omega > 0$.
Inserting Eq. \eqref{Eqdeomega} and using $I_{in}(\omega) = \frac{c}{4\pi^2}|\tilde{\mathcal{E}}(\omega)|^2$~\cite{Jackson} into Eq. \eqref{EqCrSecDeff} leads to
\begin{equation}
\sigma(\omega) = \frac{4\pi\omega}{c}\frac{\Im\left[ \tilde{d}_{\mu}(\omega) \tilde{\mathcal{E}}^{*}_{\mu}(\omega)\right]}{|\tilde{\mathcal{E}}(\omega)|^2} \,
\end{equation}
for which, we stress, a restriction to the linear regime is not invoked.

\section{Spectral resolution of the absorption cross section}\label{SecOptGapAppx}

Here, we derive the expression of Eq. \eqref{EqCrSecDeff}  for centrosymmetric systems under an incoming impulsive electric field of the form $\mathcal{E}_{\mu}(t) = \kappa_{\mu}\delta(t)$.

Let us start with the calculation of  the Fourier transform of the time-dependent dipole moment 
\begin{equation}\label{Usl2000}
d_{\mu}(t) = \theta(t)\sum\limits_{i,j=0}^{+\infty}c^*_i c_j d_{\mu}^{ij}e^{-i\omega_{ji} t}+\theta(-t)d_{\mu}^{00}.
\end{equation}
For this purpose, it is useful to consider the spectral representation of the Heaviside theta function:
\begin{equation}\label{theta1}
\theta(t) = \frac{i}{2\pi}\lim\limits_{\epsilon \to 0^+} \int\limits_{-\infty}^{+\infty} d\omega \frac{e^{-i\omega t}}{\omega+i\epsilon}.
\end{equation}
Straightforward but tedious steps lead us to
\begin{equation}\label{UUsl5}
\tilde{d}_{\mu}(\omega) = i\lim\limits_{\epsilon \to 0^+}\sum\limits_{i,j=0}^{+\infty}c_i^*c_j d_{\mu}^{ij} \left( \frac{1}{\omega-\omega_{ji}+i\epsilon}- \frac{1}{\omega-i\epsilon} \right).
\end{equation}

Next, let us evaluate $E_{exc}(\omega)$ as defined in Eq. \eqref{Eqdeomega}, using Eq. \eqref{UUsl5} together with $\tilde{\mathcal{E}}_{\mu}(\omega) = \kappa_{\mu}$.  By means of additional straightforward steps, we arrive at
\begin{equation}\label{UUsl7}
E_{exc}(\omega)= \lim\limits_{\epsilon \to 0^+}\frac{1}{\pi}\sum\limits_{i,j>i}^{+\infty}\Re\left[c_i^*c_j d_{\mu}^{ij}\kappa_{\mu} \frac{\omega_{ji}}{\omega-\omega_{ji}+i\epsilon}\\
-c_ic_j^{*} d_{\mu}^{ij}\kappa_{\mu} \frac{\omega_{ji}}{\omega+\omega_{ji}+i\epsilon} \right].
\end{equation}
We note that
$
\Re(c_ic_j^*) = \Re(c_i^*c_j)
$,
$
\Im(c_ic_j^*) = -\Im(c_i^*c_j)
$,
\begin{equation}\label{Usl32}
\Re\left( \frac{1}{\omega\pm\omega_{ji} + i\epsilon} \right) = \frac{(\omega\pm\omega_{ji})}{(\omega\pm\omega_{ji})^2+\epsilon^2},
\end{equation}
and
\begin{equation}\label{Usl33}
\Im\left( \frac{1}{\omega\pm\omega_{ji} + i\epsilon} \right) = -\frac{\epsilon}{(\omega\pm\omega_{ji})^2+\epsilon^2}.
\end{equation}
As a result, we get
\begin{multline}\label{Usl35}
E_{exc}(\omega) =\frac{1}{\pi}\lim\limits_{\epsilon \to 0^+}  \sum\limits_{i,j>i}
\left[
\Re(c_i^*c_j) d_{\mu}^{ij}\kappa_{\mu}\omega_{ji} \left(\frac{\omega-\omega_{ji}}{(\omega-\omega_{ji})^2+\epsilon^2}-\frac{\omega+\omega_{ji}}{(\omega+\omega_{ji})^2+\epsilon^2}\right)\right.\\
+\left.\Im(c_i^*c_j) d_{\mu}^{ij}\kappa_{\mu}\omega_{ji} \left(\frac{\epsilon}{(\omega-\omega_{ji})^2+\epsilon^2}-\frac{\epsilon}{(\omega+\omega_{ji})^2+\epsilon^2}\right)\right].
\end{multline}

In Eq. \eqref{Usl35}, we can use
\begin{equation}
\lim\limits_{\epsilon\to 0^+} \frac{1}{\pi}\frac{\epsilon}{(\omega_{ji}\pm\omega)^2+\epsilon^2} = \delta(\omega_{ji}\pm\omega),
\end{equation}
where $\omega_{ji} > 0$ for $j>i$. Thus, we end up with
\begin{multline}\label{Eabsotot}
E_{exc}(\omega) =  \sum\limits_{i,j>i}
\left[\lim\limits_{\epsilon \to 0^+}
\Re(c_i^*c_j) d_{\mu}^{ij}\kappa_{\mu} \left(\frac{\omega_{ji}(\omega-\omega_{ji})}{(\omega-\omega_{ji})^2+\epsilon^2}-\frac{\omega_{ji}(\omega+\omega_{ji})}{(\omega+\omega_{ji})^2+\epsilon^2}\right)\right.\\
+\left.\Im(c_i^*c_j) d_{\mu}^{ij}\kappa_{\mu}\omega_{ji} \delta(\omega_{ji}-\omega)\right] \, ,
\end{multline}
where, as clarified in the derivation of Eq. \eqref{Eqdeomega}, we restrict ourselves to $\omega > 0$.
 
Next, let us consider the case of centrosymmetric systems. In this case,
the eigenstates are either of \textit{gerade} or \textit{ungerade} parity under inversion of the coordinates.
Solely on the based of  symmetry considerations, we readily conclude that $\Re(c_i^*c_j) = 0$.
Using the expression of the $c_i$ coefficients in Eq. \eqref{WFcoeff} in the main text, we find that
\begin{multline}\label{ImCoeff}
\Im(c^*_i c_j) = \mel{\Psi_0}{\cdk}{\Psi_i}\mel{\Psi_j}{\sin(\dk )}{\Psi_0}\\-
\mel{\Psi_0}{\sdk}{\Psi_i}\mel{\Psi_j}{\cdk}{\Psi_0}.
\end{multline}
Finally, using Eq. \eqref{Eabsotot}, Eq. \eqref{ImCoeff}, and $I_{in}(\omega) = c/4\pi^2|\kappa|^2$ in Eq. \eqref{Eqdeomega}, we arrive at
\begin{multline}\label{sigmao}
\sigma(\omega) = \frac{E_{exc}(\omega)}{I_{in}(\omega)} = \frac{4\pi^2}{c|\kappa|^2}\sum\limits_{i,j>i}
\left[\cdkij{0}{i}\sdkij{j}{0}\right.\\
-\left.\sdkij{0}{i}\cdkij{j}{0}\right] \dkij{i}{j}\omega_{ji} \delta(\omega_{ji}-\omega).
\end{multline}

\section{Fifth order correction to the cross section}\label{Secsigma5Appx}

Adopting the notation $d_{\mu}^{ij} = \mel{\Psi_i}{\hat{d}_{\mu}}{\Psi_j}$ and introducing $\dkijn{i}{j}{n} = \mel{\Psi_i}{(\dk)^n}{\Psi_j}$, 
the fifth order correction to the cross section is obtained by power expansion of Eq. \eqref{CrossSectionEf} in the main text as
\begin{multline}
\sigma^{(5)}(\omega) = \frac{4\pi^2}{c\kappa^2}
\left\lbrace 
\frac{1}{120}\sum\limits_{j}
\dkijn{j}{0}{5}
\dkij{0}{j}\omega_{j0}\delta(\omega-\omega_{j0})
\right.\\
+\frac{1}{12}\sum\limits_{j > i}
\left[
  \dkijn{0}{i}{2}\dkijn{j}{0}{3} - \dkijn{0}{i}{3}\dkijn{j}{0}{2}
\right]
\dkij{i}{j}\omega_{ji}\delta(\omega-\omega_{ji})
\\
+\left.\frac{1}{24}\sum\limits_{j > i}
\left[
  \dkijn{0}{i}{4}\dkij{j}{0} - \dkij{0}{i}\dkijn{j}{0}{4}
\right]
\dkij{i}{j}\omega_{ji}\delta(\omega-\omega_{ji})
\right\rbrace\ ,
\end{multline}
{\color{blue} which can be expressed as the sum of ground and excited state absorption as
\begin{equation}
\sigma^{(5)}(\omega) = \frac{4\pi^2}{c\kappa^2} \left[ \sigma^{(5)}_{\rm GSA}(\omega) + \sigma^{(5)}_{\rm ESA}(\omega) \right] \, ,
\end{equation}
where
\begin{multline}
\sigma^{(5)}_{\rm GSA}(\omega) = \frac{1}{120}\sum\limits_{j}
\left[\dkij{0}{j}\dkijn{j}{0}{5}
+10  \dkijn{0}{0}{2}\dkij{0}{j}\dkijn{j}{0}{3} \right. \\
\left. -10  \dkijn{0}{0}{3}\dkij{0}{j}\dkijn{j}{0}{2}
+5   \dkijn{0}{0}{4}|\dkij{j}{0}|^2
\right]\omega_{j0}\delta(\omega-\omega_{j0})
\end{multline}
and
\begin{multline}
\sigma^{(5)}_{\rm ESA}(\omega) = 
\frac{1}{24}\sum_{\substack{i > 0\\j > i}}
\left\lbrace 2\left[\dkijn{0}{i}{2}\dkijn{j}{0}{3}-\dkijn{0}{i}{3}\dkijn{j}{0}{2}\right]  \right. \\
\left. +\left[\dkijn{0}{i}{4}\dkij{j}{0}-\dkij{0}{i}\dkijn{j}{0}{4}\right] \right\rbrace
\dkij{i}{j}\omega_{ji}\delta(\omega-\omega_{ji}) \ .
\end{multline}
}
\bibliography{Bibliography}